%% file: 0.paper.tex
\documentclass[jair,twoside,11pt,theapa]{article}
\usepackage{jair, theapa, rawfonts}

\usepackage{amsmath}
\usepackage{amsthm}
\usepackage{booktabs}
\usepackage{algorithm}
\usepackage[noend]{algpseudocode}
\usepackage{multirow}
\usepackage{graphicx}
\usepackage{mwe}
\usepackage{subfig}
\usepackage{csquotes}
\usepackage{amssymb}

\jairheading{81}{2024}{359-399}{07/2023}{10/2024}
\ShortHeadings{Inverting Cryptographic Hash Functions}
{Zaikin}
\firstpageno{359}

\theoremstyle{definition}
\newtheorem{example}{Example}[section]

\newcommand{\ca}{{\sim}}

\algrenewcommand\algorithmicrequire{\textbf{Input:}}
\algrenewcommand\algorithmicensure{\textbf{Output:}}
\algnewcommand\algorithmicforeach{\textbf{for each}}
\algdef{S}[FOR]{ForEach}[1]{\algorithmicforeach\ #1\ \algorithmicdo}

\begin{document}

\title{Inverting Cryptographic Hash Functions via Cube-and-Conquer}

\author{\name Oleg Zaikin \email oleg.zaikin@icc.ru \\
        \addr Novosibirsk State University \\
        Novosibirsk, 630090, Russia \\
        Matrosov Institute for System Dynamics and Control Theory SB RAS  \\ Irkutsk, 664033, Russia \\
       }

% For research notes, remove the comment character in the line below.
% \researchnote

\maketitle

\begin{abstract}
MD4 and MD5 are fundamental cryptographic hash functions proposed in the early 1990s. MD4 consists of 48 steps and produces a 128-bit hash given a message of arbitrary finite size. MD5 is a more secure 64-step extension of MD4. Both MD4 and MD5 are vulnerable to practical collision attacks, yet it is still not realistic to invert them, i.e., to find a message given a hash. In 2007, the 39-step version of MD4 was inverted by reducing to SAT and applying a CDCL solver along with the so-called Dobbertin's constraints. As for MD5, in 2012 its 28-step version was inverted via a CDCL solver for one specified hash without adding any extra constraints. In this study, Cube-and-Conquer (a combination of CDCL and lookahead) is applied to invert step-reduced versions of MD4 and MD5. For this purpose, two algorithms are proposed. The first one generates inverse problems for MD4 by gradually modifying the Dobbertin's constraints. The second algorithm tries the cubing phase of Cube-and-Conquer with different cutoff thresholds to find the one with the minimum runtime estimate of the conquer phase. This algorithm operates in two modes: (i) estimating the hardness of a given propositional Boolean formula; (ii) incomplete SAT solving of a given satisfiable propositional Boolean formula. While the first algorithm is focused on inverting step-reduced MD4, the second one is not area-specific and is therefore applicable to a variety of classes of hard SAT instances. In this study, 40-, 41-, 42-, and 43-step MD4 are inverted for the first time via the first algorithm and the estimating mode of the second algorithm. Also, 28-step MD5 is inverted for four hashes via the incomplete SAT solving mode of the second algorithm. For three hashes out of them, it is done for the first time.
\end{abstract}

\input{1.intro}
\input{2.prelim}
\input{3.algdobb}
\input{4.algcnc}
\input{5.problems}
\input{6.exper-md4-40}
\input{7.exper-md4-43}
\input{8.exper-md5}
\input{9.related}
\input{10.conc}

\acks{This study was financially supported by the Mathematical Center in Akademgorodok under the agreement No. 075-15-2022-282 with the Ministry of Science and Higher Education of the Russian Federation. The author thanks anonymous reviewers for valuable and thorough comments. The author is grateful to Stepan Kochemazov and Alexander Semenov for fruitful discussions.}
\input{appendix}

\vskip 0.2in
\bibliography{refs}
\bibliographystyle{theapa}

\end{document}

%% file: 1.intro.tex
\section{Introduction}

A cryptographic hash function maps a message of arbitrary finite size to a hash of fixed size. Consider the following properties: (i) preimage resistance; (ii) second preimage resistance; (iii) collision resistance~\cite{MenezesOV96-Handbook}. The first property indicates that it is computationally infeasible to invert the cryptographic hash function, i.e., to find any message that matches a given hash. According to the second property, given a message and its hash, it is computationally infeasible to find another message with the same hash. The third property indicates that it is computationally infeasible to find two different messages with the same hash. A proper cryptographic hash function must have all three properties. Cryptographic hash functions are ubiquitous in the modern digital world. Examples of their applications include the verification of data integrity, passwords, and signatures.

It is well known that the resistance of a cryptographic hash function can be analyzed by algorithms for solving the Boolean satisfiability problem (SAT)~\cite{Bard09-AlgCrypt}. SAT in its decision form is to determine whether a given propositional Boolean formula is satisfiable or not~\cite{BiereHMW21-SAT}. This is one of the most well-studied NP-complete problems~\cite{Cook71-NPcompl,GareyJ79-NP}. Over the last 25 years, numerous crucial scientific and industrial problems have been successfully solved by SAT. In almost all these cases, CDCL solvers, i.e., those based on the Conflict-Driven Clause Learning algorithm~\cite{Marques-SilvaS99-CDCL}, were used.

Cube-and-Conquer is an approach for solving extremely hard SAT instances~\cite{HeuleKWB11-CnC} for which CDCL solvers alone are not enough. According to this approach, a given problem is split into subproblems in the cubing phase via a lookahead solver~\cite{HeuleM2121-LookaheadHandbook}. In the conquer phase, the subproblems are solved via a CDCL solver until a satisfying assignment is found. Several hard mathematical problems from number theory and combinatorial geometry have been solved by Cube-and-Conquer recently, e.g., the Boolean Pythagorean Triples problem~\cite{HeuleKM16-BPT}. However, the authors of this study are not aware of any successful application of this approach to cryptanalysis problems. This study aims to fill this gap by analyzing the preimage resistance of the cryptographic hash functions MD4 and MD5 via Cube-and-Conquer.

MD4 was proposed in 1990~\cite{Rivest90-MD4}. It consists of 48 steps and produces a 128-bit hash given a message of arbitrary finite size. In 1995, it was shown that MD4 is not collision resistant~\cite{Dobbertin96-MD4collisions}. Since MD4 still remains preimage resistant and second preimage resistant, its step-reduced versions have been studied in this context recently. In 1998, the Dobbertin's constraints for intermediate states of MD4 registers were proposed, which reduce the number of preimages, but at the same time significantly simplify the inversion~\cite{Dobbertin98-NotOneWay}. This breakthrough approach made it possible to easily invert 32-step MD4. In 2007, SAT encodings of slightly modified Dobbertin's constraints were constructed, and as a result 39-step MD4 was inverted via a CDCL solver~\cite{DeKV07-MD4} for one very regular hash (128 1s). When the preimage resistance is studied, it is a common practice to invert very regular hashes such as all 1s or all 0s. Since 2007, several unsuccessful attempts have been made to invert 40-step MD4.

MD5 is a more secure 64-step version of MD4, which was proposed in 1992~\cite{Rivest92-MD5}. Thanks to their elegant yet efficient designs, MD4 and MD5 have become one of the most influential cryptographic functions with several notable successors, such as RIPEMD and SHA-1. Since 2005, MD5 has been known to be not collision resistant~\cite{WangY05-md5}. Because of the more secure design, the Dobbertin's constraints are not applicable to MD5~\cite{AokiS08-63stepMD5}. 26-step MD5 was inverted in 2007~\cite{DeKV07-MD4}, while for 27- and 28-step MD5 it was done for the first time in 2012~\cite{LegendreDK12-Ictai}. In both papers, CDCL solvers were applied, but no extra constraints were added. In~\cite{LegendreDK12-Ictai}, 28-step MD5 was inverted for only one hash $\mathtt{0x01234567\,0x89abcdef\,0xfedcba98\,0x76543210}$. This hash is a regular binary sequence (it is symmetric, and the numbers of 0s and 1s are equal), but at the same time it is less regular than 128 1s mentioned above. The same result was presented later in two papers by the same authors. Unfortunately, none of these three papers explained the non-existence of results for 128 1s and 128 0s. Since 2012, no further progress in inverting step-reduced MD5 has been made.

This paper proposes Dobbertin-like constraints as a generalization of Dobbertin's constraints. In addition, two algorithms are proposed. The first one generates Dobbertin-like constraints until preimages of a step-reduced MD4 are found by a complete algorithm. In the paper, SAT solving algorithms are applied for this purpose. The second algorithm does sampling to find a cutoff threshold for the cubing phase of Cube-and-Conquer with the minimum runtime estimate of the conquer phase. The algorithm operates in two modes: (i) estimating the hardness of a given formula; (ii) incomplete SAT solving of a given formula. Since the estimating mode is general, it can be applied to arbitrary SAT instances including unsatisfiable ones, while the incomplete SAT solving mode is oriented only on satisfiable SAT instances, preferably with many solutions.

Using the first algorithm and the estimating mode of the second algorithm, 40-, 41-, 42-, and 43-step MD4 are inverted for four hashes: 128 1s, 128 0s, the one from~\cite{LegendreDK12-Ictai}, and a random hash. When the best cutoff threshold and the corresponding runtime estimate are found in the estimating mode, in the conquer phase cubes are produced, and all corresponding subproblems are solved via a CDCL SAT solver. This differs from the typical conquer phase of Cube-and-Conquer, which stops when a satisfying assignment of any subproblem is found. It was done (i) to compare the total real runtime of all subproblems with the estimated runtime and (ii) to investigate how many preimages exist in the considered inverse problems.

It does not make sense to apply Dobbertin's constraints to MD5 because of its more secure design compared to MD4. Therefore, the first algorithm is not applicable to inverse problems for a step-reduced MD5. The estimating mode of the second algorithm is not well suited for inverting step-reduced MD5 because for an unconstrained inverse problem the cubing phase produces subproblems that are too hard. In this study, only the incomplete SAT solving mode of the second algorithm is applied to MD5. In particular, 28-step MD5 is inverted for the same four hashes as for MD4. All the experiments are run on a personal computer.

In summary, the contributions of the paper are:
\begin{itemize}
\item Dobbertin-like constraints as a generalization of Dobbertin's constraints.
\item An algorithm that generates Dobbertin-like constraints and the corresponding inverse problems to find preimages of a step-reduced MD4.
\item A general algorithm for finding a cutoff threshold with the minimum runtime estimate of the conquer phase of Cube-and-Conquer.
\item For the first time, 40-, 41-, 42-, and 43-step MD4 are inverted.
\item For the first time, 28-step MD5 is inverted for the two most regular hashes (128 1s and 128 0s) and a random non-regular hash.
\end{itemize}

The paper is organized as follows. Preliminaries on SAT, cryptographic hash functions, and Dobbertin's constraints are given in Section~\ref{sec:prelim}. Section~\ref{sec:alg-dobb} proposes Dobbertin-like constraints and the algorithm aimed at inverting step-reduced MD4. The Cube-and-Conquer-based algorithm is proposed in Section~\ref{sec:alg-cnc}. The considered inverse problems for step-reduced MD4 and MD5, as well as their SAT encodings, are described in Section~\ref{sec:problems}. Experimental results on inverting step-reduced MD4 and MD5 are presented in Sections~\ref{sec:invert-md4-40}, \ref{sec:invert-md4-43}, and~\ref{sec:invert-md5}. Section~\ref{sec:related} outlines related work. Finally, conclusions are drawn.

This paper builds on an earlier work~\cite{Zaikin20-MD4}, but extends it significantly in several directions. First, the algorithm for generating Dobbertin-like constraints for MD4 is improved by discarding impossible values of the last bits in the modified constraint. As a result, in most cases the considered step-reduced versions of MD4 are inverted approximately two times faster than in~\cite{Zaikin20-MD4}. Second, the incomplete SAT solving mode of the Cube-and-Conquer-based algorithm is proposed. Third, all considered step-reduced versions of MD4 are inverted for four hashes compared to two hashes in~\cite{Zaikin20-MD4}. Finally, 28-step MD5 is inverted, while in~\cite{Zaikin20-MD4} only step-reduced MD4 was studied.

%% file: 2.prelim.tex
\section{Preliminaries}\label{sec:prelim}
This section gives preliminaries on SAT, Cube-and-Conquer, cryptographic hash functions, MD4, Dobbertin's constraints, and MD5.

\subsection{Boolean Satisfiability}\label{subsec:prelim-sat}

% SAT
\textit{Boolean satisfiability problem} (SAT)~\cite{BiereHMW21-SAT} is to determine whether a given propositional Boolean formula is satisfiable or not. A formula is \textit{satisfiable} if there exists a truth assignment that satisfies it; otherwise it is \textit{unsatisfiable}. SAT is historically the first NP-complete problem~\cite{Cook71-NPcompl}. A propositional Boolean formula is in \textit{Conjunctive Normal Form} (CNF), if it is a conjunction of clauses. A \textit{clause} is a disjunction of literals, where a \textit{literal} is a Boolean variable or its negation. 

% DPLL 
The \textit{Davis–Putnam–Logemann–Loveland} (DPLL) algorithm is a complete backtracking SAT solving algorithm~\cite{DavisLL62-DPLL}. It forms a decision tree, where each internal node corresponds to a decision variable, while edges correspond to variables' values. When a decision variable is assigned, \textit{Unit Propagation} (UP) reduces the tree~\cite{DowlingG84-UP}. UP iteratively applies the \textit{unit clause rule}: if there is only one remaining unassigned literal in a clause and all other literals are assigned to False, then the literal is assigned to True. If an unsatisfied clause is encountered, a \textit{conflict} is declared, and chronological backtracking is performed.

% Lookahead
Lookahead is another complete SAT solving algorithm~\cite{HeuleM2121-LookaheadHandbook}. It improves DPLL by the following heuristic. When a decision variable should be chosen, each unassigned variable is assigned to True followed by UP, the reduction is measured, and the same is done for False assignment. \textit{Failed literal} denotes a literal for which a conflict is found during UP. If both literals of a variable are failed, the unsatisfiability of the formula is proven. If there is exactly one failed literal $l$ for some variable, then $l$ is assigned to False. This rule is known as \textit{failed literal elimination}. If for a variable both literals are not failed, the reduction measure for this variable is calculated as a combination of literal-measures. A variable with the largest reduction measure is picked as a decision variable. Thus lookahead allows one to choose good decision variables and simplifies the formula by the described reasoning. Lookahead SAT solvers are strong on random k-SAT formulae.
%In a decision tree, produced by lookahead, internal nodes correspond to decision variables. There are two types of leaves in such a tree: satisfying assignments and \textit{refuted leaves}. The latter correspond to failed literals.

% CDCL
In contrast to DPLL, in \textit{Conflict-Driven Clause Learning} (CDCL), when a conflict occurs, the reason is found and \textit{non-chronological backtracking} is performed~\cite{Marques-SilvaS99-CDCL}. To forbid the conflict, a \textit{conflict clause} is formed and added to the formula. Conflict clauses are used to limit the exploration of the decision tree and to choose proper decision variables. This complete algorithm is much more efficient than DPLL. Also, it is stronger than lookahead on non-random instances. That is why most modern complete SAT solvers are based on CDCL. Recently, CDCL was improved by local search~\cite{CaiZFB22-CDCLplusSLS}. Although local search algorithms are incomplete because they cannot prove unsatisfiability, the mentioned combination of CDCL and local search is complete.

% SAT-based cryptanalysis
Problems from the following areas can be efficiently reduced to SAT: model checking, planning, cryptanalysis, combinatorics, and bioinformatics~\cite{BiereHMW21-SAT}. When a cryptanalysis problem is reduced to SAT and solved by SAT solvers, it is called \textit{SAT-based cryptanalysis} or \textit{logical cryptanalysis}~\cite{CookM96-SATcrypt,MassacciM00-LogCrypt}. This is a special type of algebraic cryptanalysis~\cite{Bard09-AlgCrypt}. In the last two decades, SAT-based cryptanalysis has been successfully applied to stream ciphers, block ciphers, and cryptographic hash functions.

\subsection{Cube-and-Conquer}\label{subsec:prelim-cnc}

If a given SAT instance is too hard for a sequential SAT solver, it makes sense to solve it in parallel~\cite{BalyoS18-ParlellSAT}. If only complete algorithms are considered, there are two main approaches to parallel SAT solving: \textit{portfolio}~\cite{HamadiJS09-ManySAT} and \textit{divide-and-conquer}~\cite{BohmS96-ParallelSAT}. According to the portfolio approach, many different sequential SAT solvers (or maybe different configurations of the same solver) solve the same problem simultaneously. In the divide-and-conquer approach, the problem is decomposed into a family of simpler subproblems that are solved by sequential solvers.

\textit{Cube-and-conquer}~\cite{HeuleKWB11-CnC,HeuleKB18-HandbookCnC} is a divide-and-conquer SAT solving approach that combines lookahead with CDCL. In the \textit{cubing phase}, a modified lookahead solver splits a given formula into \textit{cubes}. In the \textit{conquer phase}, by joining each cube with the original formula a subformula is formed. Finally a CDCL solver is run on the subformulas. If the original formula is unsatisfiable, then all the subformulas are unsatisfiable. Otherwise, at least one subformula is satisfiable. If a satisfying assignment of any subformula is found, the conquer phase stops. Since cubes can be processed independently, the conquer phase can be easily parallelized. In~\cite{HeuleKWB11-CnC} it was proposed to process the subformulas via an incremental CDCL solver.

As mentioned in the previous subsection, lookahead is a complete algorithm. When used in the cubing phase of Cube-and-Conquer, a lookahead solver is forced to cut off some branches, thus producing cubes. Therefore, such a solver produces a decision tree, where leaves are either \textit{refuted} ones (with no possible solutions) or cubes. There are two main cutoff heuristics that decide when a branch becomes a cube. In the first one, a branch is cut off after a given number of decisions~\cite{HyvarinenJN10-CnC}. According to the second one, it occurs when the number of variables in the corresponding subproblem drops below a given threshold~\cite{HeuleKWB11-CnC}. In the present study, the second cutoff heuristic is used because it usually shows better results on hard instances~\cite{HeuleKB18-HandbookCnC}.

\subsection{Cryptographic Hash Functions}\label{subsec:prelim-hash}

A hash function $h$ is a function with the following properties~\cite{MenezesOV96-Handbook}.
\begin{enumerate}
\item \textit{Compression}: $h$ maps an input $x$ of arbitrary finite size to an output $h(x)$ of fixed size.
\item \textit{Ease of computation}: for any given input $x$, $h(x)$ is easy to compute.
\end{enumerate}

An unkeyed cryptographic hash function $h$ is a hash function that has the following properties~\cite{MenezesOV96-Handbook}.
\begin{enumerate}
\item \textit{Collision resistance}: it is computationally infeasible to find any two inputs $x$ and $x'$ such that $x \neq x', h(x) = h(x')$.
\item \textit{Preimage resistance}: for any given output $y$, it is computationally infeasible to find any of its preimages, i.e., any such input $x'$ that $h(x')=y$.
\item \textit{Second-preimage resistance}: for any given input $x$, it is computationally infeasible to find $x'$ such that $x' \neq x, h(x)=h(x')$.
\end{enumerate}

Inputs of cryptographic hash functions are usually called \textit{messages}, while outputs are called \textit{hash values} or just \textit{hashes}. Hereinafter only unkeyed cryptographic hash functions are considered.

%The first two properties are obligatory, while the remaining three are potential and can be compromised. Usually at the time of publishing a cryptographic hash function all the five properties are valid. However, subsequent publications may propose methods for abolishing one (or even all) of the three potential properties.

Methods for disproving the mentioned three properties are called \textit{collision attacks}, \textit{preimage attacks}, and \textit{second preimage attacks}, respectively. If an attack is computationally feasible, then it is called \textit{practical}. Usually it is much easier to propose a practical collision attack than practical attacks of two other types. 
%It is clear that collision resistance implies second-preimage resistance, while second-preimage attack implies a collision attack. 
This study is focused on practical preimage attacks on step-reduced cryptographic hash functions MD4 and MD5. In the rest of the paper, a practical inversion of a cryptographic hash function implies a practical preimage attack and vise versa.

\subsection{MD4}\label{subsec:prelim-md4}

The Message Digest 4 (MD4) cryptographic hash function was proposed by Ronald Rivest in 1990~\cite{Rivest90-MD4}. Given a message of arbitrary finite size, \textit{padding} is applied to obtain a message that can be divided into 512-bit blocks. Then a 128-bit hash is produced by iteratively applying the MD4 compression function to the blocks in accordance to the Merkle-Damgard construction~\cite{Merkle89,Damgard89a}.

Consider the compression function in more detail. Given a 512-bit input, it produces a 128-bit output. The function consists of three rounds, sixteen steps each, and operates by transforming data in four 32-bit registers $A, B, C, D$. If a message block is the first one, then the registers are initialized with the following constants, respectively: $\texttt{0x67452301}$; $\texttt{0xefcdab89}$; $\texttt{0x98badcfe}$; $\texttt{0x10325476}$. Otherwise, the registers are initialized with an output produced by the compression function on the previous message block. The message block $M$ is divided into sixteen 32-bit words. In every step, one register is updated by mixing one message word with the values of all four registers and an additive constant. This transformation is partially performed by a round-specific function. Additive constants are also round-specific. As a result, every round all sixteen words take part in such updates. Finally, registers are incremented by the values they had after the current block initialization, and the output is produced as a concatenation of $A, B, C, D$. The round functions and additive constants are presented in Table~\ref{tab:md4params}.

\begin{table}[ht]
{
\caption{Characteristics of MD4 rounds.}
\label{tab:md4params}}
\centering
\begin{tabular}{l c c}
\toprule
    Round & Round function & Additive constant \\
    \midrule
    1 & $F(x,y,z) = (x \wedge y) \vee (\neg x \wedge z)$ & $\texttt{0x0}$ \\
    2 & $G(x,y,z) = (x \wedge y) \vee (x \wedge z) \vee (y \wedge z)$ & $\texttt{0x5a827999}$ \\
    3 & $H(x,y,z) = x \oplus y \oplus z$ & $\texttt{0x6ed9eba1}$ \\
\bottomrule
\end{tabular}
\end{table}

Algorithm~\ref{alg:md4-compress} presents the compression function when it processes the first message block. The function $[abcd\,\,k\,\,s]_F$ stands for $a = (a + F(b,c,d) + M[k]) \lll s$, where $\lll s$ is the circular shifting to the left by $s$ bits position, while $+$ is the addition modulo $2^{32}$. The functions $[abcd\,\,k\,\,s]_G$ and $[abcd\,\,k\,\,s]_H$ stand for $a = (a + G(b,c,d) + M[k] + \texttt{0x5a827999}) \lll s$ and $a = (a + H(b,c,d) + X[k] + \texttt{0x6ed9eba1}) \lll s$, respectively.

\begin{algorithm}[ht]
\caption{MD4 compression function on the first 512-bit message block.}
\label{alg:md4-compress}
\begin{algorithmic}
\Require 512-bit message block $M$.
\Ensure Updated values of registers $A$, $B$, $C$, $D$.
\State $AA \gets A \gets \texttt{0x67452301}$
\State $BB \gets B \gets \texttt{0xefcdab89}$
\State $CC \gets C \gets \texttt{0x98badcfe}$
\State $DD \gets D \gets \texttt{0x10325476}$
\State [ABCD \,\,\,0  3]$_F$ [DABC  \,\,\,1  7]$_F$ [CDAB \,\,\,2 11]$_F$ [BCDA \,\,\,3 19]$_F$ \Comment{Steps 1-4}
\State [ABCD \,\,\,4  3]$_F$ [DABC  \,\,\,5  7]$_F$ [CDAB \,\,\,6 11]$_F$ [BCDA \,\,\,7 19]$_F$ \Comment{Steps 5-8}
\State [ABCD \,\,\,8  3]$_F$ [DABC  \,\,\,9  7]$_F$ [CDAB 10 11]$_F$ [BCDA 11 19]$_F$ \Comment{Steps 9-12}
\State [ABCD 12  3]$_F$      [DABC 13  7]$_F$       [CDAB 14 11]$_F$ [BCDA 15 19]$_F$ \Comment{Steps 13-16}
\State [ABCD  \,\,\,0  3]$_G$  [DABC  \,\,\,4  5]$_G$  [CDAB  \,\,\,8  \,\,\,9]$_G$  [BCDA 12 13]$_G$ \Comment{Steps 17-20}
\State [ABCD  \,\,\,1  3]$_G$  [DABC  \,\,\,5  5]$_G$  [CDAB  \,\,\,9  \,\,\,9]$_G$  [BCDA 13 13]$_G$ \Comment{Steps 21-24}
\State [ABCD  \,\,\,2  3]$_G$  [DABC  \,\,\,6  5]$_G$  [CDAB 10  \,\,\,9]$_G$  [BCDA 14 13]$_G$ \Comment{Steps 25-28}
\State [ABCD  \,\,\,3  3]$_G$  [DABC  \,\,\,7  5]$_G$  [CDAB 11  \,\,\,9]$_G$  [BCDA 15 13]$_G$ \Comment{Steps 29-32}
\State [ABCD  \,\,\,0  3]$_H$  [DABC  \,\,\,8  9]$_H$  [CDAB  \,\,4 11]$_H$  [BCDA 12 15]$_H$ \Comment{Steps 33-36}
\State [ABCD  \,\,\,2  3]$_H$  [DABC 10  9]$_H$  [CDAB  \,\,6 11]$_H$  [BCDA 14 15]$_H$ \Comment{Steps 37-40}
\State [ABCD  \,\,\,1  3]$_H$  [DABC  \,\,\,9  9]$_H$  [CDAB  \,\,5 11]$_H$  [BCDA 13 15]$_H$ \Comment{Steps 41-44}
\State [ABCD  \,\,\,3  3]$_H$  [DABC 11  9]$_H$  [CDAB  \,\,7 11]$_H$  [BCDA 15 15]$_H$ \Comment{Steps 45-48}
\State $A \gets A + AA$\Comment{Increment A by the initial value}
\State $B \gets B + BB$\Comment{Increment B by the initial value}
\State $C \gets C + CC$\Comment{Increment C by the initial value}
\State $D \gets D + DD$\Comment{Increment D by the initial value}
\end{algorithmic}
\end{algorithm}

In 1995, a practical collision attack on MD4 was proposed~\cite{Dobbertin96-MD4collisions}. In 2005, it was theoretically shown that on a very small fraction of messages MD4 is not  second preimage resistant~\cite{WangLFCY05-MD4}. In 2008, a theoretical preimage attack on MD4 was proposed \cite{Leurent08-MD4}. 
%Despite the found vulnerabilities, MD4 is still used to compute password-derived hashes in some operating systems of the Windows family, including Windows 10, due to backwards compatibility issues.

%
\subsection{Dobbertin's Constraints for MD4}\label{subsec:prelim-dobb}

Since MD4 is still preimage resistant and second preimage resistant from the practical point of view, its step-reduced versions have been studied recently. In 1998, Hans Dobbertin introduced additional constraints for MD4~\cite{Dobbertin98-NotOneWay}. Consider a constant 32-bit word $K$ and 32-step MD4. The constraints are as follows: $A=K$ in steps 13, 17, 21, 25; $D=K$ in steps 14, 18, 22, 26; $C=K$ in steps 15, 19, 23, 27 (numbering from 1). Further in the present paper these constraints are called \textit{Dobbertin's constraints}. Algorithm~\ref{alg:md4-compress-dobb} shows how the MD4 compression function is changed when Dobbertin's constraints are applied.

\begin{algorithm}[ht]
\caption{MD4 compression function on the first 512-bit message block with applied Dobbertin's constraints.}
\label{alg:md4-compress-dobb}
\begin{algorithmic}
\Require 512-bit message block $M$, constant word $K$.
\Ensure Updated values of registers $A$, $B$, $C$, $D$.
\State Initialize $A, B, C, D$ as in Algorithm~\ref{alg:md4-compress}
\State Steps 1--12 as in Algorithm~\ref{alg:md4-compress}
\State [ABCD 12  \,\,\,3]$_F$ \,\,\,$\textbf{A} \gets \textbf{K}$ \Comment{Constrained step 13}
\State [DABC 13  \,\,\,7]$_F$ \,\,\,$\textbf{D} \gets \textbf{K}$ \Comment{Constrained step 14} 
\State [CDAB 14 11]$_F$ \,\,\,$\textbf{C} \gets \textbf{K}$ \Comment{Constrained step 15} 
\State [BCDA 15 19]$_F$ \Comment{Step 16}
\State [ABCD  \,\,\,0  \,\,\,3]$_G$ \,\,\,$\textbf{A} \gets \textbf{K}$ \Comment{Constrained step 17}
\State [DABC  \,\,\,4  \,\,\,5]$_G$ \,\,\,$\textbf{D} \gets \textbf{K}$ \Comment{Constrained step 18}
\State [CDAB  \,\,\,8  \,\,\,9]$_G$ \,\,\,$\textbf{C} \gets \textbf{K}$ \Comment{Constrained step 19}
\State [BCDA 12 13]$_G$ \Comment{Step 20}
\State [ABCD  \,\,\,1  \,\,\,3]$_G$ \,\,\,$\textbf{A} \gets \textbf{K}$ \Comment{Constrained step 21}
\State [DABC  \,\,\,5  \,\,\,5]$_G$ \,\,\,$\textbf{D} \gets \textbf{K}$ \Comment{Constrained step 22}
\State [CDAB  \,\,\,9  \,\,\,9]$_G$ \,\,\,$\textbf{C} \gets \textbf{K}$ \Comment{Constrained step 23}
\State [BCDA 13 13]$_G$ \Comment{Step 24}
\State [ABCD  \,\,\,2  \,\,\,3]$_G$ \,\,\,$\textbf{A} \gets \textbf{K}$ \Comment{Constrained step 25}
\State [DABC  \,\,\,6  \,\,\,5]$_G$ \,\,\,$\textbf{D} \gets \textbf{K}$ \Comment{Constrained step 26}
\State [CDAB 10  \,\,\,9]$_G$ \,\,\,$\textbf{C} \gets \textbf{K}$ \Comment{Constrained step 27}
\State [BCDA 14 13]$_G$ \Comment{Step 28}
\State Steps 29--48 as in Algorithm~\ref{alg:md4-compress}
\State Increment $A, B, C, D$ as in Algorithm~\ref{alg:md4-compress}
\end{algorithmic}
\end{algorithm}

Consider step 17. $A=C=D=K$ due to constrained steps 13, 14, and 15, while $B$ is unknown. Since $G$ is the majority function, $G(x,y,y)=y$ for arbitrary $x$ and $y$. Therefore, we have

\begin{equation*}
\begin{aligned}
K = (A + G(B,C,D) + M[0] + \texttt{0x5a827999}) \lll 3 = \\
(K + G(B,K,K) + M[0] + \texttt{0x5a827999}) \lll 3 = \\
(K + K + M[0] + \texttt{0x5a827999}) \lll 3.
\end{aligned}
\end{equation*}

Then it follows

\begin{equation*}
\begin{aligned}
K \ggg 3 = 2K + M[0] + \texttt{0x5a827999},
\end{aligned}
\end{equation*}

and finally

\begin{equation*}
\begin{aligned}
M[0] = (K \ggg 3) - 2K - \texttt{0x5a827999}.
\end{aligned}
\end{equation*}

Here $-$ stands for subtraction modulo $2^{32}$. For example, if $K=\texttt{0xffffffff}$, then 

\begin{equation*}
\begin{aligned}
M[0] = \texttt{0xffffffff} - 2 \cdot \texttt{0xffffffff} - \texttt{0x5a827999} = \\
-\texttt{0xffffffff} - \texttt{0x5a827999} = \texttt{0xa57d8668}.
\end{aligned}
\end{equation*}

Thus, if $A$ is equal to a constant word in step 17, $M[0]$ becomes a constant as well. The same holds for $M[4]$, $M[8]$, $M[1]$, $M[5]$, $M[9]$, $M[2]$, $M[6]$, and $M[10]$ due to constrained steps 18, 19, 21, 22, 23, 25, 26, and 27, respectively. Finally, Dobbertin's constraints turn 9 message words out of 16 into constants. Therefore, the constrained compression function maps $\{0,1\}^{224}$ onto $\{0,1\}^{128}$ while the original one maps $\{0,1\}^{512}$ onto $\{0,1\}^{128}$. As a result, for any given hash and a randomly chosen $K$, the number of preimages (messages) is significantly reduced, maybe even to 0. Dobbertin's constraints are an example of streamlined constraints~\cite{GomesS04-Streamlined}. Such constraints are not implied by the formula, so they can remove some (or even all) solutions but have a good chance of leaving at least one solution. 

Dobbertin's constraints were originally proposed for 32-step MD4, and they do not guarantee that for a certain pair (hash,$K$) at least one preimage remains. On the other hand, they guarantee that the corresponding system of equations becomes much smaller and easier to solve. It is clear that different $K$ can be tried until a preimage is found. The point is that even if a few such simplified problems are to be solved, it may be faster than solving the original problem. The same holds for more than 32 steps because the constraints are applied before the 32nd step. In other words, adding more unconstrained steps does not reduce the number of solutions.

In~\cite{Dobbertin98-NotOneWay}, Dobbertin's constraints were used to invert 32-step MD4 by randomly choosing values of $K$ and $B$ (on step 28) until a consistent system was formed and a preimage was found. In the case of 32 steps, a constant value $B$ in addition to $K$ in step 28 implies values of the remaining 7 message words. This is not the case for more than 32 steps. In 2000, modified Dobbertin's constraints were applied to invert MD4 when the second round is omitted~\cite{KuwakadoT00-DobbExt}. In 2007, a SAT-based implementation of slightly modified Dobbertin's constraints (where the constraint in step 13 is omitted) made it possible to invert 39-step MD4~\cite{DeKV07-MD4}. Since 2007, several unsuccessful attempts to invert 40-step MD4 have been made, see, e.g.,~\cite{LegendreDK12-Ictai}. The present study aims to invert 40-, 41-, 42-, and 43-step MD4.

\subsection{MD5}\label{subsec:prelim-md5}

MD5 was proposed in 1992 by Ronald Rivest as a slightly slower but more secure extension of MD4~\cite{Rivest92-MD5}.

The main changes in MD5 compared to MD4 are as follows.
\begin{enumerate}
\item The fourth 16-step round with its own round function, so MD5 consists of 64 steps;
\item New function for the second round;
\item Usage of an unique additive constant in each of the 64 steps;
\item Addition of output from the previous step.
\end{enumerate}

%\begin{algorithm}[ht]
%\caption{One step of MD5.}
%\label{alg:md5step}
%\textbf{Input}: Current registers' values $a, b, c, d$, where $b$ is a value updated in the previous step; message word index $p$;  shift amount $s$; additive constant's index $q$; a nonlinear function $Func$.\\
%\textbf{Output}: Updated value $a$.
%\begin{algorithmic}[1]
%\State $x \gets a + Func(b,c,d) + M[p] + K[q]$
%\State $x \gets x \lll s$
%\State $a \gets b + x$
%\end{algorithmic}
%\end{algorithm}

The round functions are as follows:
\begin{itemize}
    \item Round 1. $F(x,y,z) = (x \wedge y) \vee (\neg x \wedge z)$.
    \item Round 2. $G(x,y,z) = (x \wedge z) \vee (y \wedge \neg z)$.
    \item Round 3. $H(x,y,z) = x \oplus y \oplus z$.
    \item Round 4. $I(x,y,z) = y \oplus (x \vee \neg z)$.
\end{itemize}

For the first time, a practical collision attack on MD5 was presented in 2005~\cite{WangY05-md5}. In 2009, a theoretical preimage attack was proposed~\cite{SasakiA09-md5}. It is known that Dobbertin's constraints are not efficient for MD5 because of changes 2-4 mentioned above~\cite{AokiS08-63stepMD5}. In particular, when applying to MD5, these constraints remove all solutions (preimages), so simplicity of the obtained simplified problem does not help. In 2007, 26-step MD5 was inverted~\cite{DeKV07-MD4} while in 2012, it was done for 27-, and 28-step MD5~\cite{LegendreDK12-Ictai}. In both papers, CDCL solvers were applied, yet no additional constraints were added to the corresponding formulas. 

Despite the described vulnerabilities, MD5 is still widely used to verify data integrity on operating systems of the Linux family\footnote{https://linux.die.net/man/1/md5sum}. 

%% file: 3.algdobb.tex
\section{Dobbertin-like Constraints for Inverting Step-reduced MD4}\label{sec:alg-dobb}

As mentioned in the previous section, the progress in inverting step-reduced MD4 was mainly due to Dobbertin's constraints. This section proposes Dobbertin-like constraints as their generalization. In addition, an algorithm for inverting step-reduced MD4 via Dobbertin-like constraints is proposed.

\subsection{Dobbertin-like Constraints}\label{subsec:dobb-like-constr}

Suppose that given a constant word $K$, only 11 of 12 Dobbertin's constraints hold as usual, while in the remaining constraint only $b, 0 \leq b \leq 32$ bits of the register are equal to the corresponding $b$ bits of $K$. At the same time, the remaining $32-b$ bits in the register take the opposite values to those in $K$. We denote these constraints as \textit{Dobbertin-like constraints}. It is clear that Dobbertin's constraints are a special case of Dobbertin-like constraints when $b=32$.

We denote an inverse problem for step-reduced MD4 with applied Dobbertin-like constraints as $\texttt{MD4inversion}(y,s,K,p,L)$, where
\begin{itemize}
\item $y$ is a given 128-bit hash;
\item $s$ is the number of MD4 steps (starting from the first one);
\item $K$ is a 32-bit constant word used in Dobbertin's constraints;
\item $p \in \{13, 14, 15, 17, 18, 19, 21, 22, 23, 25, 26, 27\}$ is a specially constrained step;
\item $L$ is a 32-bit word such that if $L_i=0, 0 \leq i \leq 31$, then $i$-th bit of the register's value modified in step $p$ is equal to $K_i$. Otherwise, it is equal to $\ca K_i$.
\end{itemize}

In other words, the 32-bit word $L$ serves as a bit mask and controls how similar the specially constrained register and $K$ are to each other. To make this definition clearer, three examples are given below. Hereinafter $\texttt{0hash}$ and $\texttt{1hash}$ mean 128 0s and 128 1s (i.e., 4 words $\texttt{0x00000000}$ and $\texttt{0xffffffff}$, respectively).

\begin{example}[$\texttt{MD4inversion}(\texttt{0hash},32,\texttt{0x62c7Ec0c},21,\texttt{0x00000000})$]
The problem is to invert $\texttt{0hash}$ produced by 32-step MD4. Since $L=\texttt{0x00000000}$, in step 21 the specially constrained register's value is $K$, so all 12 Dobbertin's constraints are applied as usual with $K=\texttt{0x62c7Ec0c}$. A similar inverse problem (up to choice of $K$) was solved in~\cite{Dobbertin98-NotOneWay}.
\end{example}

\begin{example}[$\texttt{MD4inversion}(\texttt{1hash},39,\texttt{0xfff00000},12,\texttt{0xffffffff})$]
The problem is to invert $\texttt{1hash}$ produced by 39-step MD4. Since $L=\texttt{0xffffffff}$, in step 12 the specially constrained register's value is $\ca K = \texttt{0x000fffff}$. In the remaining 11 Dobbertin's steps the registers' values are $K=\texttt{0xfff00000}$.
\end{example}

\begin{example}[$\texttt{MD4inversion}(\texttt{1hash},40,\texttt{0xffffffff},12,\texttt{0x00000003})$]
The problem is to invert $\texttt{1hash}$ produced by 40-step MD4. Since $L=\texttt{0x00000003}$, in step 12 the first 30 bits of the specially constrained register are equal to those in $K$, while the last two bits have values $\ca K_{30}$ and $\ca K_{31}$, respectively. Therefore, this register's value is $\texttt{0xfffffff\textbf{c}}$, while in the remaining 11 Dobbertin's steps the registers' values are $K=\texttt{0xffffffff}$.
\end{example}

\subsection{Inversion Algorithm}\label{subsec:inv-alg}

Dobbertin-like constraints can be used to find preimages of a step-reduced MD4 according to the following idea. For a given hash $y$, step $s$, and constant $K$, an inverse problem is formed with $L=\texttt{0x00000000}$. Thus, all 12 Dobbertin's constraints are applied. The inverse problem is solved, and if a preimage is found, nothing else should be done. Otherwise, if it is proven that no preimages exist in the current inverse problem, a new one is formed with $L=\texttt{0x00000001}$. In this case, the specially constrained register's value is just 1 bit shy of being $K$. The inverse problem is also solved. If still no preimage, $L$ is further modified: $\texttt{0x00000002}, \texttt{0x00000003}$, and so on. The intuition here is that Dobbertin's constraints lead to a system of equations that is either consistent with very few solutions or quite close to a consistent one. In the latter case, trying different values of $L$ helps to form a consistent system and find its solution.

Algorithm~\ref{alg:md4-constraints} follows the described idea. In the pseudocode, the function \textsc{DecimalToBinary} converts a decimal number to binary, while $\mathcal{A}$ is a complete algorithm, which for a formed inverse problem returns preimages if they exist. In the while loop, all possible values of the specially constrained register are varied, yet the first values are very close to $K$. Note that it is not guaranteed that Algorithm~\ref{alg:md4-constraints} finds any preimage for a given hash. However, as it will be shown in sections~\ref{sec:invert-md4-40} and~\ref{sec:invert-md4-43}, Algorithm~\ref{alg:md4-constraints} is able to find preimages for step-reduced MD4. Moreover, it usually does it in just few iterations (from 1 to 3) of the while loop.

\begin{algorithm}[h]
\caption{Invert step-reduced MD4 via Dobbertin-like constraints.}
\label{alg:md4-constraints}
\begin{algorithmic}
\Require Hash $y$; the number of MD4 steps $s$; constant $K$; step $p$ with the specially constrained register; a complete algorithm $\mathcal{A}.$
\Ensure Preimages for hash $y$.
\State $\texttt{preimages} \gets \{\}$
\State $i \gets 0$
\While{$i < 2^{32}$}
\State $L \gets \Call{DecimalToBinary}{i}$
\State $\texttt{preimages} \gets \mathcal{A}(\texttt{MD4inversion}(y,s,K,p,L))$
\If{$\texttt{preimages}$ is not empty}
\State \textbf{break}
\EndIf
\State $i \gets i + 1$
\EndWhile
\State \textbf{return} $\texttt{preimages}$
\end{algorithmic}
\end{algorithm}

Complete algorithms of various types can be used to solve inverse problems formed in Algorithm~\ref{alg:md4-constraints}. In particular, wide spectrum of constraint programming~\cite{RossiBW06-HandbookCP} solvers are potential candidates. In preliminary experiments, we used state-of-the-art sequential and parallel CDCL SAT solvers~\cite{BalyoS18-ParlellSAT} to invert 40-step MD4, but even in the first iteration, where all 12 Dobbertin's constraints are added, SAT instances turned out to be too hard for them. That is why we decided to use Cube-and-Conquer SAT solvers, which are more suitable for extremely hard SAT instances~\cite{HeuleKB18-HandbookCnC}. The next section describes how a given problem can be properly split into simpler subproblems in the cubing phase of Cube-and-Conquer.

%% file: 4.algcnc.tex
\section{Finding Cutoff Thresholds for Cube-and-Conquer}\label{sec:alg-cnc}

Recall (see Subsection~\ref{subsec:prelim-cnc}) that in Cube-and-Conquer the following cutoff threshold is meant in the cubing phase: the number of variables in a subformula, formed by adding a cube to the original Boolean propositional formula and applying UP. It is crucial to properly choose this threshold. If it is too high, then the cubing phase is performed in no time, but very few extremely hard (for a CDCL solver) subformulas might be produced; if it is too low, then the cubing phase will be extremely time consuming, while there will be a huge number of subformulas.

Earlier two algorithms aimed at finding a cutoff threshold with the minimum estimated runtime of Cube-and-Conquer were proposed. Subsection 7.2 of the tutorial~\cite{Heule18-CnCtutorial} proposed Algorithm A as follows:
\begin{displayquote}
Optimizing the heuristics requires selecting
useful subproblems of the hard formula. This can be done as follows: First determine the depth for which the number of refuted nodes is at least 1000. Second, randomly pick about 100 subproblems (cubes) of the partition on that depth. Second, solve these 100 subproblems and select the 10 hardest ones for the optimization.
\end{displayquote}

Later, Algorithm B was proposed in~\cite{BrightCSKG21-Lam}:
\begin{displayquote}
The cut-off bound was experimentally chosen by randomly selecting up to several hundred instances from each case and determining a bound that minimizes the sum of the cubing and conquering times.
\end{displayquote}

This section proposes a new algorithm inspired by algorithms A and B. The algorithm aims to find a cutoff threshold with the minimum runtime estimate of the conquer phase, so the runtime of the cubing phase is not taken into account because it is assumed that the latter is negligible.

First, it is needed to preselect promising thresholds. On the one hand, the number of refuted leaves should be quite significant since it may indicate that at least some subformulas have become simpler compared to the original formula. In addition, the total number of cubes should not be too large. An auxiliary Algorithm~\ref{alg:preselect-thresholds} follows this idea. Given a lookahead solver, a CNF, and a cutoff threshold, the function \textsc{LookaheadWithCut} runs the solver on the CNF with the cutoff threshold (see Subsection~\ref{subsec:prelim-cnc}) and outputs cubes and the number of refuted leaves.

\begin{algorithm}[h]
\caption{Preselect promising thresholds for the cubing phase of Cube-and-Conquer.}
\label{alg:preselect-thresholds}
\begin{algorithmic}
\Require CNF $\mathcal{F}$; lookahead solver $\texttt{ls}$; starting threshold $n_{\texttt{start}}$; threshold decreasing step $\texttt{nstep}$; maximum number of cubes $\texttt{maxc}$; minimum number of refuted leaves $\texttt{minr}$.
\Ensure Stack of promising thresholds and corresponding cubes.
\Function{PreselectThresholds}{$\mathcal{F}$, $\texttt{ls}$, $n_{\texttt{start}}$, $\texttt{nstep}$, $\texttt{maxc}$, $\texttt{minr}$}
\State $\texttt{stack} \gets \{\}$
\State $n \gets n_{\texttt{start}}$
\While{$n > 0$}
\State $\langle c, r \rangle \gets \Call{LookaheadWithCut}{\texttt{ls},\mathcal{F},n}$\Comment{Get cubes and number of refuted.}
\If {$\Call{Size}{c} > \texttt{maxc}$}
  \State \textbf{break}\Comment{Break if too many cubes.}
\EndIf
\If {$r \geq \texttt{minr}$}
  \State $\texttt{stack}.\mathtt{push}(\langle n, c \rangle)$\Comment{Add threshold and cubes.}
\EndIf
\State $n \gets n - \texttt{nstep}$\Comment{Decrease threshold.}
\EndWhile
\State \textbf{return} $stack$
\EndFunction
\end{algorithmic}
\end{algorithm}

Note that it is intended that the found cutoff threshold will be used to solve all corresponding subproblems by a CDCL SAT solver in the conquer phase. This setting differs from the typical conquer phase of Cube-and-Conquer, which stops upon finding a satisfying assignment of any subproblem (see Subsection~\ref{subsec:prelim-cnc}). The setting was chosen (i) to compare the total real runtime of all subproblems with the estimated runtime and (ii) to investigate how many preimages exist in the considered inverse problems for step-reduced MD4. However, later in this section it will be discussed how the algorithm can be easily modified to fit the typical setting of the conquer phase.

When promising values of the threshold are preselected by, it is needed to estimate the hardness of the corresponding conquer phases. It can be done by choosing a fixed number of cubes by simple random sampling~\cite{StarnesYM2010-Statistics} among those produced in the cubing phase. If all corresponding subproblems from the sample are solved by a CDCL solver in a reasonable time, then an estimated total solving time for all subproblems can be easily calculated. This idea is implemented as Algorithm~\ref{alg:finding-cutoff}. 

\begin{algorithm}[h]
\caption{Find a cutoff threshold with the minimum estimated runtime of the conquer phase.}
\label{alg:finding-cutoff}
\begin{algorithmic}
\Require CNF $\mathcal{F}$; lookahead solver $\texttt{ls}$; threshold decreasing step $\texttt{nstep}$; maximum number of cubes $\texttt{maxc}$; minimum number of refuted leaves $\texttt{minr}$; sample size $N$; CDCL solver $\texttt{cs}$; CDCL solver time limit $\texttt{maxcst}$; number of CPU cores $\texttt{cores}$; operating mode $\texttt{mode}$.
\Ensure A threshold $n_{\texttt{best}}$ with the runtime estimate $e_{\texttt{best}}$ and cubes $c_{\texttt{best}}$; Boolean $\texttt{isSAT}$ that indicates whether $\mathcal{F}$ is satisfiable.
\State $\texttt{isSAT} \gets \texttt{Unknown}$
\State $n_{\texttt{start}} \gets \Call{Varnum}{\mathcal{F}} - \texttt{nstep}$
\State $\langle n_{\texttt{best}}, e_{\texttt{best}}, c_{\texttt{best}} \rangle \gets \langle n_{\texttt{start}}, +\infty, \{\} \rangle$
\State $\texttt{stack} \gets \Call{PreselectThresholds}{\mathcal{F}, \texttt{ls}, n_{\texttt{start}}, \texttt{nstep}, \texttt{maxc}, \texttt{minr}}$\Comment{First stage.}
\While{$\texttt{stack}$ is not empty}\Comment{Second stage: estimate thresholds.}
\State $\langle n, c \rangle \gets \texttt{stack}.\texttt{pop()}$\Comment{Get a threshold and cubes.}
\State $\texttt{sample} \gets \Call{SimpleRandomSample}{c,N}$\Comment{Select $N$ random cubes.}
\State $\texttt{runtimes} \gets \{\}$
\ForEach{$\texttt{cube}$ \textbf{from} $\texttt{sample}$}
\State $\langle t, \texttt{st} \rangle \gets \Call{SolveCube}{\texttt{cs},\mathcal{F},\texttt{cube},\texttt{maxcst}}$\Comment{Add a cube and solve.}
\If{$t \geq \texttt{maxcst}$ \textbf{and} $\texttt{mode} = \texttt{estimating}$}\Comment{If CDCL was interrupted}
    \State \textbf{break}\Comment{in estimating mode, stop processing sample.}
\Else
    \State $\texttt{runtimes.add}(t)$\Comment{Add proper runtime.}
\EndIf
\If{$st = \texttt{True}$}\Comment{If SAT,}
    \State $\texttt{isSAT} \gets \texttt{True}$
    \If{$\texttt{mode} = \texttt{incomplete-solving}$}\Comment{and incomplete SAT solving mode,}
        \State \textbf{return} $\langle n_{\texttt{best}}, e_{\texttt{best}}, c_{\texttt{best}}, \texttt{isSAT} \rangle$\Comment{return SAT immediately.}
    \EndIf
\EndIf
\EndFor
\If{$\Call{Size}{\texttt{runtimes}} < N$}\Comment{If at least one  interrupted in sample,}
    \State \textbf{break}\Comment{stop main loop.}
\EndIf
\State $e \gets \Call{Mean}{\texttt{runtimes}} \cdot \Call{Size}{c} / \texttt{cores}$\Comment{Calculate a runtime estimate.}
\If {$e < e_{\texttt{best}}$}
    \State $\langle n_{\texttt{best}}, e_{\texttt{best}}, c_{\texttt{best}} \rangle \gets \langle n, e, c \rangle$\Comment{Update best threshold.}
\EndIf
\EndWhile
\State \textbf{return} $\langle n_{\texttt{best}}, e_{\texttt{best}}, c_{\texttt{best}}, \texttt{isSAT} \rangle$
\end{algorithmic}
\end{algorithm}

In the first stage, promising thresholds are preselected by the function \textsc{PreselectThresholds} which is described in Algorithm~\ref{alg:preselect-thresholds}, while in the second stage the one with the minimum runtime estimate of the conquer phase is chosen among them. Given a CDCL solver, a CNF, a cube, and a time limit in seconds, the function \textsc{SolveCube} adds the cube to the CNF as unit clauses, runs the CDCL solver with the time limit on the formed CNF, and returns the runtime in seconds and an answer whether the CNF is satisfiable or not.

The algorithm operates in two modes. In the \textit{estimating} mode, the algorithm terminates upon reaching a time limit by the CDCL solver on any subproblem from random samples. In the \textit{incomplete SAT solving} mode~\cite{KautzSS21-HandbookIncompleteSAT}, the algorithm terminates upon finding a satisfying assignment. The first mode is aimed at estimating the hardness of a given CNF, while the second one aims to find a satisfying assignment of a satisfiable CNF.

The proposed algorithm has the following features. 
\begin{enumerate}
\item A stack is used to preselect promising cutoff thresholds in the first stage in order to start the second stage with solving the simplest subproblems. It allows obtaining some estimate quickly and then improve it.
\item In the \textit{estimating} mode, if in the second stage a CDCL solver fails solving some subproblem within the time limit, the algorithm terminates. This is done because in this case it is impossible to calculate a meaningful estimate for the threshold. Another reason is that subproblems from next thresholds will likely be even harder.
\item It is possible that satisfying assignments are found when solving subproblems from random samples. Indeed, if a given CNF is satisfiable, then cubes which imply satisfying assignments might be chosen to samples.
\item In the \textit{estimating} mode, even if a satisfying assignment is found when solving some subproblem from samples, the algorithm does not terminate because in this case the main goal is to calculate a runtime estimate. 
\item In the \textit{estimating} mode it is a general algorithm that is able to estimate the hardness of an arbitrary CNF.
\item In the \textit{incomplete SAT solving} mode, a solution can be found only for a satisfiable CNF, and even in this case this is not guaranteed because of the time limit for the CDCL solver.
\item Algorithm 5 can be easily modified to be oriented on finding only one solution in the conquer phase. For this purpose it is required to remove $\texttt{mode}$ and return SAT if $\texttt{st} = True$.
\item The proposed runtime estimation is a stochastic costly black-box objective function~\cite{AudetH17-BlackboxOpt,SemenovZK2021-handbook}. The algorithm minimizes this objective function.
\end{enumerate}

Since all details of Algorithm~\ref{alg:finding-cutoff} are given, it now can be compared with Algorithms A and B (see the beginning of this section). It is clear that the idea is the same in all three algorithms --- for a certain value of the cutoff threshold, a sample of cubes is formed, the corresponding subproblems are solved, and finally a runtime estimate is calculated. However, there are several major differences which are listed below.

\begin{enumerate}
    \item Algorithms A and B were described informally and briefly, while Algorithm~\ref{alg:finding-cutoff} is presented formally and in detail.
    \item In contrast to Algorithms A and B, Algorithm~\ref{alg:finding-cutoff} takes into account the situation when some subproblems from a sample are so hard that they can not be solved in reasonable time by a CDCL solver.
    \item Algorithms A and B assume that further in the conquer phase incremental solving is applied to subproblems~\cite{HeuleKWB11-CnC}, while Algorithm~\ref{alg:finding-cutoff} assumes that every subproblem is solved by a non-incremental CDCL solver.
\end{enumerate}

The main difference is the second one. This feature of Algorithm~\ref{alg:finding-cutoff} is extremely important in application to cryptanalysis problems, which are considered in the rest of the present paper. The reason is that in this case subproblems in a sample usually differ by thousands and even millions of times in CDCL solver's runtime. A possible explanation why this feature was not taken into account in both Algorithms A and B is that they were applied to combinatorial and geometric problems, where subproblems' hardness in a sample is usually uniform. Importance of the third feature follows from the second one~--- incremental SAT solving is efficient in the case of the uniform hardness, otherwise it can significantly slow down the solving process.

When a cutoff threshold is found by Algorithm~\ref{alg:finding-cutoff}, the conquer phase operates as follows. First, subformulas are created by adding cubes to the original CNF in the form of unit clauses. Second, all subformulas are solved by the same CDCL solver that was used to find the threshold for the cubing phase. In contrast to Algorithm~\ref{alg:finding-cutoff}, here the CDCL solver's runtime is not limited. Finally, the conquer phase outputs a list of all found satisfying assignments (if any) or UNSAT if all subproblems turned out to be UNSAT.

%\begin{algorithm}[tb]
%\caption{Finding all solutions in the conquer phase}
%\label{alg:cnc-solve}
%\textbf{Input}: CNF $\mathcal{F}$, CDCL solver $S$, cubes $cbs$

%\textbf{Output}: result (SAT or UNSAT) and a set of %satisfying assignments $total\_solutions$.
%\begin{algorithmic}[1]
%\State $unsat\_cubes \gets 0$
%\State $total\_solutions \gets \emptyset$
%\WHILE{condition}
%\State Do some action.
%\ENDWHILE
%\State \textbf{return} $\langle \lambda_{best}, g_{best} \rangle$
%\end{algorithmic}
%\end{algorithm}

%\subsection{Algorithm for Estimating Hardness of Extremely Hard SAT Problems}\label{subsec:alg-estim}

%The next section describes certain studied inversion problems for MD4 and MD5 as well as their SAT encodings.

%% file: 5.problems.tex
\section{Considered Inverse Problems and Their SAT Encodings}\label{sec:problems}

This section describes the considered inverse problems for step-reduced MD4 and MD5, as well as their SAT encodings.

Following all earlier attempts to invert step-reduced MD4 via SAT~\cite{DeKV07-MD4,LegendreDK12-Ictai,LafitteNH14-MD4,GribanovaS18-MD4}, the padding is omitted (see Subsection~\ref{subsec:prelim-md4}) and only one 512-bit message block is considered. Therefore, a step-reduced MD4 compression function is considered when it operates on the first block, like it was shown in Algorithm~\ref{alg:md4-compress}. The final incrementing is also omitted since it should be done only after 48-th step. Note that these restrictions does not make inverse problems easier since the compression function is the main component of MD4 function from the resistance point view. Inversion of step-reduced MD5 compression function is considered in similar way. For the sake of simplicity, from now on, MD4 means the MD4 compression function, and the same for MD5. 

\subsection{Considered Hashes}\label{subsec:enc-hashes}

The following four hashes are chosen for inversion: 
\begin{enumerate}
\item $\texttt{0x00000000\,0x00000000\,0x00000000\,0x00000000}$;
\item $\texttt{0xffffffff\,0xffffffff\,0xffffffff\,0xffffffff}$;
\item $\texttt{0x01234567\,0x89abcdef\,0xfedcba98\,0x76543210}$;
\item $\texttt{0x62c7Ec0c\,0x751e497c\,0xd49a54c1\,0x2b76cff8}$.
\end{enumerate}

Recall that $\texttt{0hash}$ and $\texttt{1hash}$ mean the first and the second hash from the list, respectively. These two hashes are chosen for inversion because it is a common practice in the cryptographic community. The reason is that inverting a hash that looks like a random word is suspicious. Indeed, one can take a random message, produce its hash and declare that this very hash is inverted. On the other hand, if a hash has a regular structure, this approach does not work. All 0s and all 1s are two hashes with the most regular structure, that is why they are usually chosen. For the first time 32-step MD4 was inverted for $\texttt{0hash}$~\cite{Dobbertin98-NotOneWay}, while in 39-step case it was done for \texttt{1hash}~\cite{DeKV07-MD4}, and later for $\texttt{0hash}$~\cite{LegendreDK12-Ictai}. As for the cryptographic hash functions SHA-0 and SHA-1, their 23-step (out of 80) versions for the first time were inverted for $\texttt{0hash}$~\cite{LegendreDK12-Ictai}.

The third hash from the list was used to invert 28-step MD5 in~\cite{LegendreDK12-Ictai}. Hereinafter this hash is called $\texttt{symmhash}$. It is symmetrical~--- the last 64 bits are the first 64 bits in reverse order, but at the same time it is less regular than $\texttt{1hash}$ or $\texttt{0hash}$. The same result for 28-step MD5 was described in two more papers of the same authors. Unfortunately, none of these three papers explain why $\texttt{1hash}$ and $\texttt{0hash}$ were not considered.

The fourth hash from the list is chosen randomly. The goal is to show that the proposed approach is applicable not only to hashes with regular structure. This hash is further called $\texttt{randhash}$.

\subsection{Step-reduced MD4}\label{subsec:enc-md4}

In contrast to~\cite{DeKV07-MD4}, where only $K=\texttt{0x00000000}$ was used as a constant in Dobbertin's constraints, in the present study $K=\texttt{0x00000000}$ and $K=\texttt{0xffffffff}$ are tried in Dobbertin-like constraints. The constraint in step 12 is chosen for the modification (so $p=12$, see Subsection~\ref{subsec:dobb-like-constr}) since in~\cite{DeKV07-MD4} the constraint for this very step was entirely omitted. Eight step-reduced versions of MD4 from 40 to 47 steps, as well as the full MD4 are studied. Hence there are $9 \times 4 \times 2 = 72$ MD4-related inverse problems in total. None of these 72 inverse problems have been solved so far.
 
Consider 40-step MD4. Since $K$ has two values and $y$ has four values, Algorithm~\ref{alg:md4-constraints} should be run on eight inputs. As a result, according to the notation from Subsection~\ref{subsec:dobb-like-constr}, the following eight inverse problems are formed in the corresponding first iterations of Algorithm~\ref{alg:md4-constraints}:
\begin{enumerate}
\item $\texttt{MD4inversion}(\texttt{0hash},40,\texttt{0x00000000},12,\texttt{0x00000000})$;
\item $\texttt{MD4inversion}(\texttt{0hash},40,\texttt{0xffffffff},12,\texttt{0x00000000})$;
\item $\texttt{MD4inversion}(\texttt{1hash},40,\texttt{0x00000000},12,\texttt{0x00000000})$;
\item $\texttt{MD4inversion}(\texttt{1hash},40,\texttt{0xffffffff},12,\texttt{0x00000000})$.
\item $\texttt{MD4inversion}(\texttt{symmhash},40,\texttt{0x00000000},12,\texttt{0x00000000})$;
\item $\texttt{MD4inversion}(\texttt{symmhash},40,\texttt{0xffffffff},12,\texttt{0x00000000})$;
\item $\texttt{MD4inversion}(\texttt{randhash},40,\texttt{0x00000000},12,\texttt{0x00000000})$;
\item $\texttt{MD4inversion}(\texttt{randhash},40,\texttt{0xffffffff},12,\texttt{0x00000000})$.
\end{enumerate}

For illustrative purpose, consider the first case: invert $\texttt{0hash}$ produced by 40-step MD4 with Dobbertin's constraints and $K=\texttt{0x00000000}$. If no preimage exists for this inverse problem, then in the second iteration of Algorithm~\ref{alg:md4-constraints} $L$ is increased by 1, so the inverse problem $\texttt{MD4inversion}(\texttt{0hash},40,\texttt{0x00000000},12,\texttt{0x00000001})$ is formed and so on.

\subsection{Step-reduced MD5}\label{subsec:enc-md5}

In the present paper, inversion of 28-step MD5 compression function is considered for the four hashes presented above. Recall that in contrast to MD4, no extra constraints that reduce the number of preimages are added. Note that for all hashes but \texttt{symmhash} the inverse problems have not been solved earlier.

\subsection{SAT Encodings}\label{subsec:enc}

It is possible to construct CNFs that encode MD4 and MD5 by the following automatic tools: \textsc{CBMC}~\cite{ClarkeKL2004-CBMC}; \textsc{SAW}~\cite{CarterFHHT13-SAW}; \textsc{Transalg}~~\cite{SemenovOG0K20-LMCS}; \textsc{CryptoSAT}~\cite{Lafitte18-CryptoSAT}. In the present paper, the CNFs are constructed via \textsc{Transalg}\footnote{https://gitlab.com/transalg/transalg} of version 1.1.5. This tool takes a description of an algorithm as an input and outputs a CNF that implements the algorithm. The description must be formulated in a domain specific C-like language called TA language. TA language supports the following basic constructions used in procedural languages: variable declarations; assignment operators; conditional operators; loops; function calls. In addition it supports various integer operations and bit operations including bit shifting and comparison that is quite handy when describing a cryptographic algorithm. A TA program is a list of functions in TA language. All the constructed CNFs and the corresponding TA programs are available online as a dataset~\cite{Zaikin2024-ZenodoJAIR}. All these CNFs can be easily reconstructed by giving the TA programs to \textsc{Transalg} as inputs.

In a CNF that encodes step-reduced MD4, the first 512 variables correspond to a message, the last 128 variables correspond to a hash, while the remaining auxiliary variables are needed to encode how the hash is produced given the message. The first 512 variables are further called \textit{message variables}, while the last 128 ones --- \textit{hash variables}. The Tseitin transformations are used in \textsc{Transalg} to introduce auxiliary variables~\cite{Tseitin70}.

Characteristics of the constructed CNFs are given in Table~\ref{tab:enc}. 

\begin{table}[ht]
{
\caption{Characteristics of CNFs that encode the considered step-reduced MD4 and MD5.}
\label{tab:enc}}
\centering
\begin{tabular}{l c c c}
\toprule
    Function & Variables & Clauses & Literals \\
    \midrule
    MD4-40 & 7025 & 70~809 & 317~307 \\
    MD4-41 & 7211 & 73~158 & 329~330\\
    MD4-42 & 7397 & 75~507 & 341~353 \\
    MD4-43 & 7583 & 77~856 & 353~376 \\
    MD4-44 & 7769 & 80~205 & 365~399 \\
    MD4-45 & 7955 & 82~554 & 377~422 \\
    MD4-46 & 8141 & 84~903 & 389~445 \\
    MD4-47 & 8327 & 87~252 & 401~468 \\
    MD4-48 & 8513 & 89~601 & 413~491 \\
    \midrule
    MD5-28 & 7471 & 54~672 & 216~362 \\
\bottomrule
\end{tabular}
\end{table}

The CNF that encodes 40-step MD4 has 7~025 variables and 70~809 clauses. Then every step adds 186 variables and 2~349 clauses, so as a result the CNF that encodes the full (48-step) MD4 has 8~513 variables and 89~601 clauses. Note that these CNFs encode the functions themselves, so all message and hash variables are unassigned. To obtain a CNF that encodes an inverse problem for a given 128-bit hash, corresponding 128 one-literal clauses are to be added, so all hash variables become assigned. The problem is to find values of the message variables. Dobbertin's constraints are added as another 384 unit clauses~--- 32 clauses for every constraint. As a result, the CNF that encodes the inversion of 40-step MD4 with all 12 Dobbertin's constraints has 7~025 variables and 71~321 clauses, while that for the 48-step version consists of 8~513 variables and 90~113 clauses. Note that Dobbertin-like constraints (see Subsection~\ref{subsec:dobb-like-constr}) are also added as 384 unit clauses; the only difference is in values of the corresponding 32 variables that encode the specially constrained register.

The CNF that encodes 28-step MD5 has 7~471 variables and 54~672 clauses. A CNF that encodes an inverse problem has 7~471 variables and 54~800 clauses since only 128 unit clauses for hash variables are added.

%% file: 6.exper-md4-40.tex
\section{Inverting 40-step MD4 via Dobbertin-like Constraints}\label{sec:invert-md4-40}

This section first discusses how the proposed algorithms can be combined. Then the experimental setup, simplification, and results for 40-step MD4 are described.

\subsection{Possible Combinations of Proposed Algorithms}

Recall that in every iteration of Algorithm~\ref{alg:md4-constraints} an inverse problem is formed by varying values of $L$ and is solved by some complete algorithm $\mathcal{A}$. Assume that a step-reduced MD4 is to be inverted given a hash. Then the following combinations of Algorithm~\ref{alg:md4-constraints} and the estimating mode of Algorithm~\ref{alg:finding-cutoff} are possible:
\begin{enumerate}
    \item In Algorithm~\ref{alg:md4-constraints}, $\mathcal{A}$ is (i) encoding an inverse problem to a CNF, (ii) finding the best cutoff threshold for the CNF by the estimating mode of Algorithm~\ref{alg:finding-cutoff}, and (iii) running the conquer phase of Cube-and-Conquer with the found cutoff threshold on the CNF.
    \item A CNF is formed for $L=\texttt{0x00000000}$ as in the first iteration of Algorithm~\ref{alg:md4-constraints}, and the best cutoff threshold for it is found by the estimating mode of Algorithm~\ref{alg:finding-cutoff}. Then Algorithm~\ref{alg:md4-constraints} starts such that $\mathcal{A}$ is (i) encoding an inverse problem to a CNF and (ii) running the conquer phase of Cube-and-Conquer with the found cutoff threshold on the CNF.
    %For each hash, its own best threshold is found for $L=\texttt{0x00000000}$ and is used for all other values of $L$. In Algorithm~\ref{alg:md4-constraints}, $\mathcal{A}$ is again a Cube-and-Conquer solver with the found threshold.
    %\item The estimating mode of Algorithm~\ref{alg:finding-cutoff} is run on a CNF that encodes the inversion problem for an arbitrary hash among given ones, yet the Dobbertin's constraints are fully applied, i.e. $L=\texttt{0x00000000}$. When the best cutoff threshold is found, Algorithm~\ref{alg:md4-constraints} is iteratively run using a Cube-and-Conquer solver with the found threshold as algorithm $\mathcal{A}$ on all given hashes. It means that the threshold found for one hash and $L=\texttt{0x00000000}$ is used for all other hashes and values of $L$.
\end{enumerate}

In the first combination, its own cutoff threshold is found in every iteration, while in the second one, a cutoff threshold is found once and then it is used in every iteration. In this study, the second combination is used to reduce the number of Algorithm~\ref{alg:finding-cutoff} calls. Note that for any value of $L$ the same amount of unit clauses is added to a CNF.

\subsection{Experimental Setup}\label{subsec:exp-setup}

Algorithm~\ref{alg:md4-constraints} was implemented in Python, while Algorithm~\ref{alg:finding-cutoff} and the conquer phase of Cube-and-Conquer were implemented in C++ as a parallel SAT solver \textsc{Estimate-and-Cube-and-Conquer} (EnCnC). The sources are available at github\footnote{https://github.com/olegzaikin/EnCnC}, while the sources, benchmarks, and solutions are available at Zenodo~\cite{Zaikin2024-ZenodoJAIR}.

All experiments were executed on a personal computer equipped with the 12-core CPU AMD 3900X. The implementations are multithreaded, so all 12 CPU cores were employed. In case of Algorithm~\ref{alg:finding-cutoff}, values of a cutoff threshold and then subproblems from samples are processed in parallel. In case of the conquer phase, subproblems are processed in parallel.

The inputs of Algorithm~\ref{alg:md4-constraints} in case of 40-step MD4 were discussed in Section~\ref{sec:problems}. As for Algorithm~\ref{alg:finding-cutoff}, the inputs are as follows:

\begin{itemize}
\item \textsc{March\_cu} lookahead solver~\cite{HeuleKWB11-CnC} since it has been recently successfully applied to several hard problems~\cite{HeuleKM16-BPT,Heule18-Schur5}.
\item $\texttt{nstep}=5$. It was chosen in preliminary experiments. If this parameter is equal to 1, then a better threshold usually can be found, but at the same time Algorithm~\ref{alg:finding-cutoff} becomes quite time consuming. On the other hand, if $nstep$ is quite large, e.g., 50, then as a rule almost all most promising thresholds are skipped.
\item $\texttt{maxc} = 2~000~000$. On the considered CNFs, \textsc{March\_cu} reaches 2~000~000 cubes in about 30 minutes, so this value of $maxc$ looks reasonable. Higher values were also tried, but it did not give any improvement.
\item $\texttt{minr} = 1~000$. If it is less then 1~000, then subproblems are too hard because they are not simplified enough by lookahead. At the same time, higher value of this parameter did not allow collecting enough amount of promising thresholds.
\item $N=1~000$. First $N=100$ was tried, but it led to too optimistic estimates which were several times lower than real solving time. On the other hand, $N=10~000$ is too time consuming and gives just modest improvement in accuracy compared to $N=1~000$. The accuracy of obtained estimates is discussed later in Subsection~\ref{subsec:discussion}.
\item \textsc{Kissat} CDCL solver of version sc2021~\cite{BiereFH21-kissat} because this solver and its modifications won the SAT Competition in 2020--2023.
\item $\texttt{maxst}=5~000$ seconds. It is a standard time limit in SAT Competitions~\cite{BalyoFHIJS-2021-SatCompet}, so modern CDCL solvers are designed to show all their power within this time.
\item $\texttt{cores}=12$.
\item $\texttt{mode}=\texttt{estimating}$. 
\end{itemize}

%Here the goal is not just to find one preimage, but rather find all preimages for a given inverse problem (up to added Dobbertin-like constraints).

%It should be noted that in both Algorithm~\ref{alg:finding-cutoff} and the conquer phase of Cube-and-Conquer subproblems were solved by \textsc{Kissat} in the non-incremental mode, i.e., it solved them independently from each other.

%
\subsection{Simplification}\label{subsec:invert-md4-40-simpl}
In case of 40-step MD4, two parameters were varied for each of four considered hashes (see Section~\ref{subsec:enc-hashes}). The first one is the value of the Dobbertin's constant $K$ (see Section~\ref{sec:alg-dobb}): $\texttt{0x00000000}$ and $\texttt{0xffffffff}$. The second one is simplification applied to a CNF. A motivation behind varying the second parameter is as follows. First, it is crucial to simplify a CNF before giving it to a lookahead solver. Second, in preliminary experiments it was found out that the simplification type can significantly alter the effectiveness of Cube-and-Conquer on the considered problems.

The CDCL solver \textsc{CaDiCaL} of version 1.5.0~\cite{FroleyksB-Cadical1.5} was used to simplify the CNFs. This solver uses inprocessing, i.e., a given CNF is simplified during the CDCL search \cite{Biere11-PreInProcessing}. The more conflicts have been generated by a CDCL solver so far, the more simplified (in terms of the number of variables) the CNF has been made. A natural simplification measure in this case is the number of generated conflicts. In the experiments related to 40-step MD4, the following limits on the number of generated conflicts were tried: 1, 10 thousand, 100 thousand, 1 million, 10 million. Note that 1 conflict as the limit in some cases gives the same result as UP (see Subsection~\ref{subsec:prelim-sat}), while in the remaining cases the corresponding CNF is slightly smaller.

For example, consider problem $\texttt{MD4inversion}(\texttt{1hash},40,\texttt{0xffffffff},12,\texttt{0x00000000})$. Table~\ref{tab:simpl} presents characteristics of six CNFs which encode this problem. The original (unsimplified) CNF is described by the number of variables, clauses, and literals. For those simplified by \textsc{CaDiCaL}, also the runtime on 1 CPU core is given.

\begin{table}[ht]
{\caption{CNFs that encode $\texttt{MD4inversion}(\texttt{1hash},40,\texttt{0xffffffff},12,\texttt{0x00000000})$.  The best values are marked with bold.}
\label{tab:simpl}}
\centering
\begin{tabular}{l c c c c}
\toprule
    Simplification type & Variables & Clauses & {Literals} & Simplification runtime \\
    \midrule
    no (original CNF) & 7025 & 71~321 & 317~819 & - \\
    1 conflict & 3824 & 33~371 & 138~820 & 0.02 sec \\
    10 thousand conflicts & 2969 & 27~355 & 116~618 & 0.31 sec \\
    100 thousand conflicts & 2803 & 23~121 & 94~250 & 4.29 sec \\
    1 million conflicts & 2756 & \textbf{22\,391} & \textbf{90\,412} & 1 min 19 sec \\
    10 million conflicts & \textbf{2054} & 24~729 & 110~267 & 33 min \\ 
\bottomrule
\end{tabular}
\end{table}

It is clear, that first the number of variables, clauses, and literals decrease, and then 10 million conflicts provides lower number of variables yet the number of clauses and literals is higher than that on 1 million conflicts. For other hashes and values of $L$ the picture is similar.

\subsection{Experiments}\label{subsec:invert-md4-40-exper}
Of course, more parameters can be varied for each hash in addition to $K$ and simplification type mentioned in the previous subsection. One of the most natural is a CDCL solver used in Cube-and-Conquer. For example, a cryptanalysis oriented solver can be chosen~\cite{SoosNC09-cm,NejatiG19-CdclCrypto,Kochemazov21-CryptSatSolver}. Moreover, internal parameters of the chosen CDCL solver can be varied as well.

Recall that there are 4 hashes and 5 simplification types, while $K$ has 2 values, so $4 \times 5 \times 2 = 40$ CNFs were constructed with fully applied Dobbertin's constraints ($L=\texttt{0x00000000}$) for MD4-40. On each of them the first iteration of Algorithm~\ref{alg:md4-constraints} was run. It turned out, that Algorithm~\ref{alg:finding-cutoff} could not find any estimates for all 20 CNFs with $K=\texttt{0x00000000}$. The reason is that in all these cases \textsc{Kissat} was interrupted due to the time limit even for the simplest (lowest) values of the cutoff threshold. On the other hand, for $K=\texttt{0xffffffff}$ much more positive results were achieved. For \texttt{0hash}, \texttt{symmhash}, and \texttt{randhash}, estimates for all simplification types were successfully calculated, and the best type was 1 conflict in all these cases. On the other hand, for \texttt{1hash} no estimates were found for 1 conflict and 10 thousand conflicts, while the best estimate was gained for 1 million conflicts.

The results are presented in Table~\ref{tab:md4-40-est}. For each pair (simplification type, hash), the best estimate for 12 CPU cores, the corresponding cutoff threshold, and the number of cubes are given. Here ``-'' means that no estimate was obtained because \textsc{Kissat} was interrupted on the simplest threshold. Runtimes of Algorithm~\ref{alg:finding-cutoff} are not presented there, but on average it took about 2 hours for $K=\texttt{0x00000000}$ and about 3 hours for $K=\texttt{0xffffffff}$.

\begin{table}[ht]
{\caption{Runtime estimates for 40-step MD4. The best estimates are marked with bold.}
\label{tab:md4-40-est}}
\centering
\begin{tabular}{c c c c c}
\toprule
    Hash & Simplification conflicts & $e_{\texttt{best}}$ & $n_{\texttt{best}}$ & $|c_{\texttt{best}}|$ \\
    \midrule
    \multirow{5}{*}{\texttt{0}} & \textbf{1} & \textbf{15 h 33 min} & \textbf{3290} & \textbf{303\,494} \\
    & 10 thousand & 21 h 43 min & 2530 & 210~008 \\
    & 100 thousand & 52 h 32 min & 2485 & 107~657 \\
    & 1 million & 22 h 19 min & 2400 & 148~518 \\
    & 10 million & 34 h 27 min & 1895 & 69~605 \\
    \midrule
    \multirow{5}{*}{\texttt{1}} & 1 & - & - & - \\
    & 10 thousand & - & - & - \\
    & 100 thousand & 81 h 31 min & 2535 & 362~429 \\
    & \textbf{1 million} & \textbf{42 h 43 min} & \textbf{2510} & \textbf{182\,724} \\
    & 10 million & 991 h 12 min & 1890 & 1~671~849 \\
    \midrule
    \multirow{5}{*}{\texttt{symm}} & \textbf{1} & \textbf{19 h 16 min} & \textbf{3395} & 80~491 \\
    & 10 thousand & 29 h 47 min & 2725 & 181~267 \\ 
    & 100 thousand & 22 h 44 min & 2615 & 60~403 \\
    & 1 million & 21 h 11 min & 2530 & 151~567 \\
    & 10 million & 59 h 28 min & 1945 & 189~744 \\
    \midrule
    \multirow{5}{*}{\texttt{rand}} & \textbf{1} & \textbf{14 h 27 min} & \textbf{3400} & \textbf{75\,823} \\
    & 10 thousand & 227 h 54 min & 2660 & 1~098~970 \\ 
    & 100 thousand & 20 h 22 min & 2540 & 159~942 \\
    & 1 million & 17 h 33 min & 2455 & 225~854 \\
    & 10 million & 81 h 3 min & 1915 & 242~700 \\
\bottomrule
\end{tabular}
\end{table}

\begin{figure}[ht]
   \centering
  \includegraphics[width=.7\textwidth]{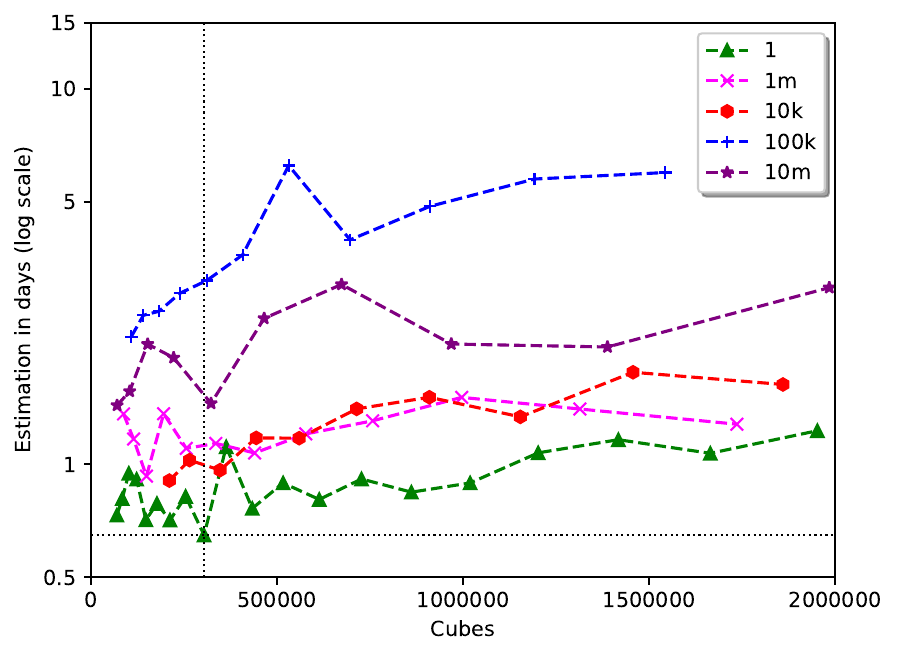}
    \caption{Minimization of the objective function on 40-step MD4, \texttt{0hash}. The intersection of two dotted lines shows the best estimate.}
    \label{fig:est-md4-40-0hash}
\end{figure}

Figure~\ref{fig:est-md4-40-0hash} depicts how the objective function was minimized on the inverse problem for \texttt{0hash}. Here 10k stands for 10 thousand conflicts, 1m for 1 million conflicts and so on. The figures for the remaining three hashes can be found in Appendix~\ref{appendix:figures}.

\begin{figure}[!ht]
   \centering
   \includegraphics[width=.7\textwidth]{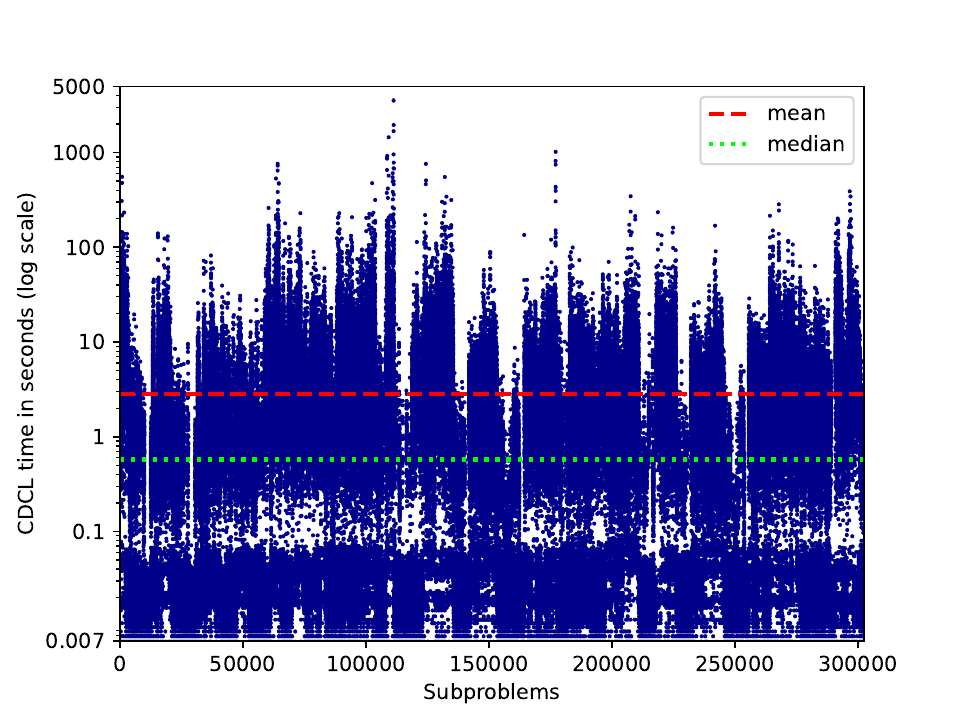}
    \caption{\textsc{Kissat} runtimes on subproblems from the conquer phase applied to $\texttt{MD4inversion}(\texttt{0hash},40,\texttt{0xffffffff},12,\texttt{0x00000000})$}.
    \label{fig:solving-md4-40-0hash}
\end{figure}

As mentioned in Section~\ref{sec:alg-cnc}, in the estimating mode of Algorithm~\ref{alg:finding-cutoff} it is possible to find satisfying assignments of a given satisfiable CNF. That is exactly what happened for \texttt{symmhash} --- a satisfying assignments was found for the CNF simplified by 100 thousand conflicts. It means that a preimage for \texttt{symmhash} generated by 40-step MD4 was found just in few hours during the search for good thresholds for the cubing phase. However, the goal was to solve all subproblems of the considered inverse problems (up to chosen value of $L$) to find more preimages. That is why using the cubes produced with the help of the best cutoff thresholds, the conquer phase was run on all four inverse problems: 1-conflict-based for \texttt{0hash}, \texttt{symmhash}, and \texttt{randhash}; 1-million-conflicts-based for \texttt{1hash}. As a result, all subproblems were solved successfully. The subproblems' runtimes in case of \texttt{0hash} are shown in Figure~\ref{fig:solving-md4-40-0hash}.

For \texttt{0hash} and \texttt{1hash}, no satisfying assignments were found, therefore the corresponding inverse problems have no solutions. On the other hand, satisfying assignments were found for $\texttt{symmhash}$ and $\texttt{randhash}$. The found thresholds, estimates, and the real runtimes are presented in Table~\ref{tab:runtimes-md4-40}. In the header, \texttt{sol} stands for the number of solutions. Note that the best estimate $e_{best}$ was calculated only for $L=\texttt{0x00000000}$, so for other values of $L$ it is equal to ``-''. The right three columns present subproblems' statistics: mean runtime; maximum runtime; and standard deviation of runtimes (when they are in seconds). The minimum runtime is not reported since it was equal to 0.007 seconds in all cases.

\begin{table}[ht]
{\caption{Estimated and real runtimes (on 12 CPU cores) of the conquer phase for inverse problems related to 40-step MD4. The best estimates from Table~\ref{tab:md4-40-est} are presented.}
\label{tab:runtimes-md4-40}}
\centering
\begin{tabular}{cccccccc}
\toprule
    Hash & $L$ & $e_{\texttt{best}}$ & real time & sol & mean & max & $\texttt{sd}$ \\
	\midrule
	\multirow{3}{*}{\texttt{0}} & $\texttt{0x00000000}$ & 15 h 33 min & 20 h 9 min & 0 & 2.84 sec & 1 h & 13.68 \\
	& $\texttt{0x00000001}$ & - & 19 h 25 min & 0 & 2.61 sec & 29 min & 10.22 \\
	& $\texttt{0x00000002}$ & - & 34 h 27 min & 1 & 5.7 sec & 38 min & 20.62 \\
	\midrule
	\multirow{3}{*}{\texttt{1}} & $\texttt{0x00000000}$ & 42 h 43 min & 48 h 29 min & 0 & 11.5 sec & 26 min & 30.23 \\
	& $\texttt{0x00000001}$ & - & 59 h 7 min & 0 & 4.08 sec & 17 min & 11.4 \\
    & $\texttt{0x00000002}$ & - & 28 h 1 min & 1 & 7.7 sec & 17 min & 18.68 \\
	\midrule
	\texttt{symm} & $\texttt{0x00000000}$ & 19 h 16 min & 20 h 45 min & 2 & 11.24 sec & 18 min & 21.1 \\
	\midrule
	\texttt{rand} & $\texttt{0x00000000}$ & 14 h 27 min & 15 h 48 min & 1 & 9.08 sec & 38 min & 21.59 \\
\bottomrule
\end{tabular}
\end{table}

The next iteration of Algorithm~\ref{alg:md4-constraints} (with $L=\texttt{0x00000001}$) was executed for \texttt{0hash} and \texttt{1hash}. Note that the same simplification and cutoff threshold as for $L=\texttt{0x00000000}$ were applied to the corresponding CNFs. The conquer phase again did not find any satisfying assignment. Finally, preimages for both hashes were found in the third iteration ($L=\texttt{0x00000002}$), see Table~\ref{tab:runtimes-md4-40}. All found preimages are presented in Table~\ref{tab:preimages-md4-40}. The obtained results will be discussed in the next section.

\begin{table}[ht]
{\caption{Found preimages for 40-step MD4.}
\label{tab:preimages-md4-40}}
\centering
\begin{tabular}{cl}
\toprule
Hash & Preimages \\
\midrule
\multirow{3}{*}{\texttt{0}} &
\scriptsize{0xe57d8668 0xa57d8668 0xa57d8668 0xbc8c857b 0xa57d8668 0xa57d8668 0xa57d8668 0xcb0a1178} \\
& \scriptsize{0xa57d8668 0xa57d8668 0xa57d8668 0x307bc4e7 0xad02e703 0xe1516b23 0x981c2a75 0xc08ea9f7} \\ 
\midrule
\multirow{3}{*}{\texttt{1}} &
\scriptsize{0xe57d8668 0xa57d8668 0xa57d8668 0x1d236482 0xa57d8668 0xa57d8668 0xa57d8668 0x97a13204} \\ 
& \scriptsize{0xa57d8668 0xa57d8668 0xa57d8668 0x991ede3 0x301e2ac3 0x5bed2a3d 0xe167a833 0x890d22f0} \\
\midrule
\multirow{4}{*}{\texttt{symm}} 
& \scriptsize{0xa57d8668 0xa57d8668 0xa57d8668 0xc8cf2f7c 0xa57d8668 0xa57d8668 0xa57d8668 0x61915bc1} \\
& \scriptsize{0xa57d8668 0xa57d8668 0xa57d8668 0x2c017cc4 0xda6acfa2 0x55e9f993 0x50d83f7b 0x2d7d47a6} \\
\cline{2-2}
& \scriptsize{0xa57d8668 0xa57d8668 0xa57d8668 0x154f3b86 0xa57d8668 0xa57d8668 0xa57d8668 0x95b7616d}\\
& \scriptsize{0xa57d8668 0xa57d8668 0xa57d8668 0xf3ca15df 0x7eb66f5e 0x446dc43f 0x7d8e2888 0xafe37a76} \\
\midrule
\multirow{2}{*}{\texttt{rand}} 
& \scriptsize{0xa57d8668 0xa57d8668 0xa57d8668 0xbb809ab0 0xa57d8668 0xa57d8668 0xa57d8668 0xab67285f} \\
& \scriptsize{0xa57d8668 0xa57d8668 0xa57d8668 0x85517639 0xc3eab3d 0x6edfba39 0xa1512693 0xaa686ac9} \\
\bottomrule
\end{tabular}
\end{table}

%% file: 7.exper-md4-43.tex
\section{Inverting 41-, 42-, and 43-step MD4 via Dobbertin-like Constraints}\label{sec:invert-md4-43}

This section presents results on inverting 41-, 42-, and 43-step MD4. Finally, all MD4-related results are discussed.

\subsection{41-, 42-, and 43-step MD4}\label{subsec:exper-41-43md4}

Recall that in the previous section on inverting 40-step MD4, Algorithm~\ref{alg:md4-constraints} was run on 40 CNFs: for each of 4 hashes, 2 values of $K$ and 5 simplification types were tried. Note that $K=\texttt{0x00000000}$ did not allow solving any 40-step-related problem. As for simplification types, for 3 hashes out of 4 the best estimates were obtained on 1-conflict-based CNFs, while for the remaining one 1 million conflicts was the best. Following these results, in this section only $K=\texttt{0xffffffff}$ and two mentioned simplification types are used. Therefore, only 8 CNFs were constructed for 41-step MD4, and the same for 42-, 43-, and 44-step MD4. Also it turned out that the best 40-step-related estimates were achieved when at most 303~494 cubes were produced, see Table~\ref{tab:md4-40-est}. That is why in this section the value of $maxc$ is reduced from $2~000~000$ to $500~000$. The remaining input parameters of Algorithm~\ref{alg:finding-cutoff} are the same.

The same approach was applied as in the previous section: for each pair (the number of steps, hash) first the best cutoff threshold was found via Algorithm~\ref{alg:finding-cutoff} for a CNF with added Dobbertin's constraints ($L=\texttt{0x00000000}$), and then Algorithm~\ref{alg:md4-constraints} used the found threshold to run the conquer phase of Cube-and-Conquer as a complete algorithm in each iteration. For 44 steps, no estimates were obtained. On the other hand, for 41, 42, and 43 steps estimates were successfully calculated and they turned out to be comparable to that for 40 steps. Moreover, Algorithm~\ref{alg:finding-cutoff} found preimages for two problems: 41 step and \texttt{1hash}; 42 steps and \texttt{0hash}. In Section~\ref{sec:alg-cnc} it was discussed that such a situation is possible if a given CNF is satisfiable. The found estimates for 43-step MD4 are presented in Table~\ref{tab:md4-43-est}. For all hashes, 1 conflict was the best. For 41 steps, 1 conflict was better on \texttt{0hash} and \texttt{1hash}, while on the remaining two hashes 1-million-conflicts based simplification was the winner. On 42-step MD4, 1 conflict was the best for all hashes except \texttt{1hash}.

\begin{table}[ht]
{\caption{Runtime estimates for 43-step MD4. The best estimates are marked with bold.}
\label{tab:md4-43-est}}
\centering
\begin{tabular}{ccccc}
\toprule
    Hash & Simplif. conflicts & $e_{\texttt{best}}$ & $n_{\texttt{best}}$ & $|c_{\texttt{best}}|$ \\
    \midrule
    \multirow{2}{*}{\texttt{0}} 
    & \textbf{1} & \textbf{15 h 26 min} & \textbf{3~390} & \textbf{103~420} \\
    & 1 million & - & - & - \\
    \midrule
    \multirow{2}{*}{\texttt{1}}
    & \textbf{1} & \textbf{39 h 10 min} & \textbf{3~395} & \textbf{98~763} \\
    & 1 million & 52 h 5 min & 2~575 & 121~969 \\
    \midrule
    \multirow{2}{*}{\texttt{symm}} 
    & \textbf{1} & \textbf{37 h 51 min} & \textbf{3~395} & \textbf{81~053} \\
    & 1 million & 50 h 7 min & 2~555 & 253~489 \\
    \midrule
    \multirow{2}{*}{\texttt{rand}} 
    & \textbf{1} & \textbf{49 h 13 min} & \textbf{3~385} & \textbf{120~619} \\
    & 1 million & 86 h 23 min & 2~565 & 246~972 \\
\bottomrule
\end{tabular}
\end{table}

\begin{figure}[!ht]
   \centering
   \includegraphics[width=.7\textwidth]{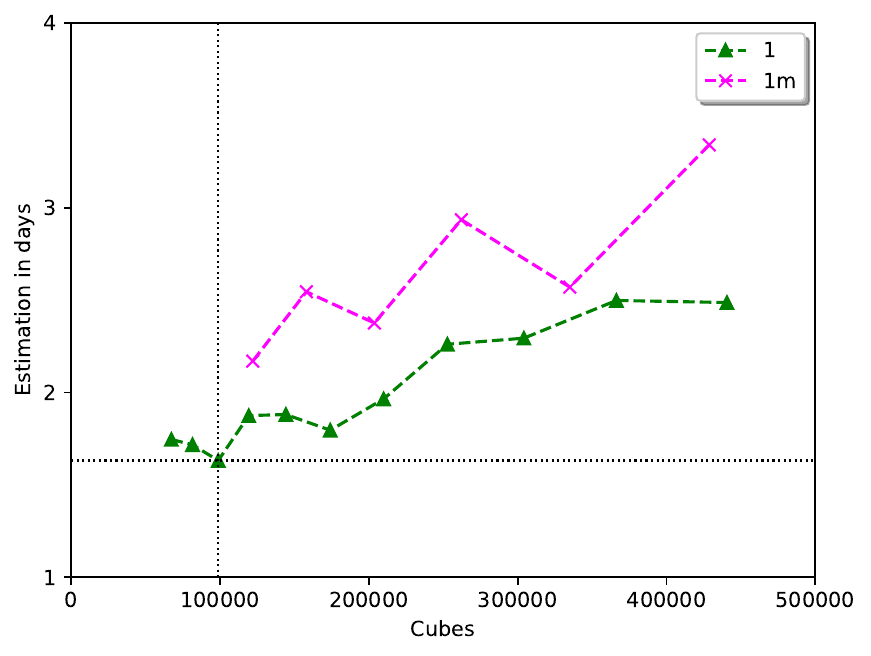}
    \caption{Minimization of the objective function on the inverse problem $\texttt{MD4inversion}(\texttt{1hash},43,\texttt{0xffffffff},12,\texttt{0x00000000})$. The intersection of two dotted lines shows the best estimate among all simplification types.}
    \label{fig:est-md4-43-1hash}
\end{figure}

Figure~\ref{fig:est-md4-43-1hash} depicts how the objective function was minimized on the inverse problem for \texttt{1hash} in case of 43 steps. Figures for the remaining three 43-steps-related inverse problems are presented in Appendix~\ref{appendix:figures}.

Using the found cutoff thresholds, Algorithm~\ref{alg:md4-constraints} was run on all inverse problems with $L=\texttt{0x00000000}$, and for 43 steps preimages were found for all four hashes. For steps 41 and 42, preimages were found in the first or the second iteration of Algorithm~\ref{alg:md4-constraints}. The results are presented in Table~\ref{tab:runtimes-md4-41-42-43}. Here values 0 and 1 of $L$ stand for $\texttt{0x00000000}$ and $\texttt{0x00000001}$, respectively, while sd stands for standard deviation in seconds. It can be seen that at least some inverse problems turned out to be easier compared to that for 40-step MD4. This phenomenon is discussed in the next subsection.

\begin{table}[h]
{\caption{Estimated and real runtimes (on 12 CPU cores) of the conquer phase for inverse problems related to 41-, 42, and 43-step MD4.}
\label{tab:runtimes-md4-41-42-43}}
\centering
\begin{tabular}{ccccccccc}
\toprule
Steps & Hash & $L$ & $e_{\texttt{best}}$ & real time & sol & mean & max & $\texttt{sd}$ \\
\midrule
\multirow{6}{*}{41} & \multirow{2}{*}{\texttt{0}} & 0 & 8 h 40 min & 10 h 11 min & 0 & 6.4 sec & 17 min & 16.77 \\ 
\cline{3-9}
& & 1 & - & 21 h 23 min & 1 & 12.41 sec & 14 h 23 min & 421.41 \\
\cline{2-9}
& \texttt{1} & 0 & 37 h & 45 h 10 min & 3 & 9.78 sec & 52 min & 44.73 \\
\cline{2-9}
& \multirow{2}{*}{\texttt{symm}} & 0 & 19 h 54 min & 20 h 10 min & 0 & 12.08 sec & 17 min & 24.28 \\
\cline{3-9}
& & 1 & - & 20 h 15 min & 4 & 11.57 sec & 17 min & 23.66 \\
\cline{2-9}
& \texttt{rand} & 0 & 16 h 6 min & 17 h 25 min & 1 & 10.05 sec & 43 min & 31.07 \\
\midrule
\multirow{6}{*}{42} & \texttt{0} & 0 & 19 h 36 min & 22 h 32 min & 3 & 11.68 sec & 19 min & 25.51 \\
\cline{2-9}
& \multirow{2}{*}{\texttt{1}} & 0 & 25 h 15 min & 29 h 19 min & 0 & 10.91 sec & 1 h 14 min & 45.61 \\
\cline{3-9}
& & 1 & - & 39 h & 1 & 16.38 sec & 2 h 18 min & 86.32 \\
\cline{2-9}
& \texttt{symm} & 0 & 28 h 20 min & 29 h 35 min & 1 & 12.25 sec & 32 min & 19.98\\
\cline{2-9}
& \multirow{2}{*}{\texttt{rand}} & 0 & 21 h 16 min & 21 h 30 min & 0 & 10.22 sec & 15 min & 18.51 \\
\cline{3-9}
& & 1 & - & 20 h 35 min & 3 & 9.34 sec & 13 min & 16.71 \\
\midrule
\multirow{4}{*}{43} & \texttt{0} & 0 & 15 h 26 min & 17 h 14 min & 2 & 7.23 sec & 16 min & 16.6 \\
\cline{2-9}
& \texttt{1} & 0 & 39 h 10 min & 42 h 16 min & 1 & 18.64 sec & 39 min & 29.88 \\
\cline{2-9}
& \texttt{symm} & 0 & 37 h 51 min & 41 h 55 min & 1 & 22.59 sec & 34 min & 46.44 \\
\cline{2-9}
& \texttt{rand} & 0 & 49 h 13 min & 51 h 21 min & 1 & 18.51 sec & 46 min & 30.41 \\
\bottomrule
\end{tabular}
\end{table}

\begin{figure}[!ht]
   \centering
   \includegraphics[width=.7\textwidth]{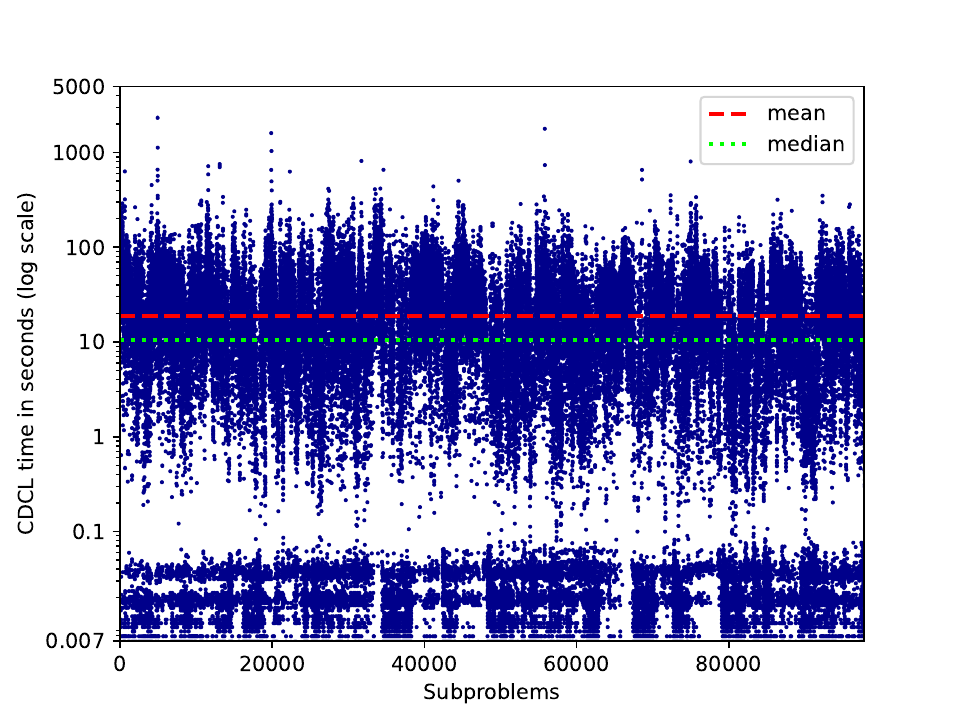}
    \caption{\textsc{Kissat} runtimes on subproblems from the conquer phase applied to $\texttt{MD4inversion}(\texttt{1hash},43,\texttt{0xffffffff},12,\texttt{0x00000000})$.}
    \label{fig:solving-md4-43-1hash}
\end{figure}

The subproblems' runtimes for 43 steps and \texttt{1hash} are shown in Figure~\ref{fig:solving-md4-43-1hash}.
In Table~\ref{tab:preimages-md4-43}, the found preimages for 43-step MD4 are presented. The corresponding tables for steps 41 and 42 are presented in Appendix~\ref{appendix:preimages}.

\begin{table}
{\caption{Found preimages for 43-step MD4.}
\label{tab:preimages-md4-43}}
\centering
\begin{tabular}{cl}
\toprule
Hash & Preimages \\
\midrule
\multirow{4}{*}{\texttt{0}}
& \scriptsize{0xa57d8668 0xa57d8668 0xa57d8668 0xf48a97a3 0xa57d8668 0xa57d8668 0xa57d8668 0xd330e8ed} \\ 
& \scriptsize{0xa57d8668 0xa57d8668 0xa57d8668 0x37c9ca21 0xe1df551f 0x7f49d66a 0x135a1c93 0x9e744bdb} \\ 
\cline{2-2}
& \scriptsize{0xa57d8668 0xa57d8668 0xa57d8668 0xb289afa0 0xa57d8668 0xa57d8668 0xa57d8668 0xaf2c850e} \\
& \scriptsize{0xa57d8668 0xa57d8668 0xa57d8668 0x19c5ce09 0xcae6b29e 0xb2595b20 0xab3a433d 0xf6cdee42} \\
\midrule
\multirow{2}{*}{$\texttt{1}$} 
& \scriptsize{0xa57d8668 0xa57d8668 0xa57d8668 0x82ef987a 0xa57d8668 0xa57d8668 0xa57d8668 0xe18fbc3b} \\
& \scriptsize{0xa57d8668 0xa57d8668 0xa57d8668 0x558f3513 0xbf09004d 0x8fb490dd 0x502eca9 0xbd0e1a80} \\
\midrule
\multirow{2}{*}{\texttt{symm}}
& \scriptsize{0xa57d8668 0xa57d8668 0xa57d8668 0xd1c33d35 0xa57d8668 0xa57d8668 0xa57d8668 0xc8519181} \\
& \scriptsize{0xa57d8668 0xa57d8668 0xa57d8668 0x8157aaf2 0xd7bdc37b 0xe52f3348 0xf17901d9 0x7e2de5a4} \\
\midrule
\multirow{2}{*}{\texttt{rand}}
& \scriptsize{0xa57d8668 0xa57d8668 0xa57d8668 0x24f0e099 0xa57d8668 0xa57d8668 0xa57d8668 0xe57e4c54} \\
& \scriptsize{0xa57d8668 0xa57d8668 0xa57d8668 0x8fbbadcd 0xc0326ae6 0xe0e6a048 0x6217a3b9 0x15ee5a3b} \\
\bottomrule
\end{tabular}
\end{table}

\subsection{Discussion}\label{subsec:discussion}

\paragraph{Correctness} 
The correctness of the found preimages was verified by the reference implementation from~\cite{Rivest90-MD4}. This verification can be easily reproduced since MD4 is hard to invert, but the direct computation is extremely fast. First, the additional actions (padding, incrementing, see Section~\ref{sec:problems}), as well as the corresponding amount of the last steps should be deleted. Then the found preimages should be given as inputs to the compression function.

\paragraph{Simplification}
According to the estimates, in most cases the 1-conflict-based simplification is better than more advanced simplifications. On the other hand, if only this simplification type had been chosen, then the inverse problem for \texttt{1hash} produced by 40-step MD4 would have remained unsolved. The non-effectiveness of the advanced simplifications is an interesting phenomenon which is worth investigating in the future.

\paragraph{Classes of subproblems}
Figures~\ref{fig:solving-md4-40-0hash} and~\ref{fig:solving-md4-43-1hash} show that in the conquer phase about 25\% of subproblems are extremely easy (runtime is less than 0.1 second) and there is a clear gap between these subproblems and the remaining ones. Since this gap is much lower than mean and median runtime, is seems promising to solve all extremely easy subproblems beforehand and apply the corresponding reasoning to the remaining subproblems.

\paragraph{Estimation accuracy}
The obtained estimates can be treated as accurate since they are close to the real solving times (see Tables~\ref{tab:runtimes-md4-40} and~\ref{tab:runtimes-md4-41-42-43}). On average the real time on inverse problems with $L=\texttt{0x00000000}$ is 11 $\%$ higher than the estimated time, while in the worst case for 40-step MD4 and \texttt{0hash} the real time is 30 $\%$ higher. As for the real time on inverse problems with $L=\texttt{0x00000001}$ and $L=\texttt{0x00000002}$, the situation is different. In some cases, the real time is still close to the estimated time for $L=\texttt{0x00000000}$. However, for $\texttt{MD4inversion}(\texttt{0hash},41,\texttt{0xffffffff},12,\texttt{0x00000001})$ the real time is 2.5 times higher, while the standard deviation is also very high. It can be concluded that the heavy-tail behavior occurs in this case~\cite{GomesS21-Heavytail}. These results might indicate that it is better to find its own cutoff threshold for each value of $L$, that corresponds to the first combination of Algorithm~\ref{alg:md4-constraints} and Algorithm~\ref{alg:finding-cutoff} described at the beginning of Section~\ref{sec:invert-md4-40}. Note that for those problems where their own thresholds were used, i.e., when $L=\texttt{0x00000000}$, the heavy-tail behavior does not occur.

\paragraph{Hardness of inverse problems}
It might seem counterintuitive that for 40-43 steps the hardness of the inverse problems in fact is more or less similar. Recall that when Dobbertin's constraints are applied, values of 9 message words of 16 with indices 0, 1, 2, 4, 5, 6, 8, 9, and 10 are derived automatically (in a CNF this is done by UP), so only 7 words remain unknown (see Subsection~\ref{subsec:prelim-dobb}). It means that in the CNF 224 message bits are unknown compared to 512 message bits when Dobbertin's constraints are not added. It holds true for Dobbertin-like constraints as well. In the 40th step, the register's value is updated via a round function that takes as input an unknown word $M[14]$ along with registers' values. That is why the 40th step gives a leap in hardness compared to 39 steps. In the next 8 steps, message words with the following indices are used for updating: $1, 9, 5, 13, 3, 11, 7, 15$. It means that in steps 41, 42, and 43 the round function operates with known (constant) $M[1]$, $M[9]$, and $M[5]$, respectively. In MD4 compression function, the main hardness is added by mixing a message word with registers' values. Therefore, steps 41-43 do not add any hardness. Rather, additional connections between registers' values are added. As for the remaining steps 44-48, only unknown message words are used for updating, so each of these steps gives a new leap in hardness. That is why no estimates were calculated for 44 steps earlier in this section~--- those inverse problems are much harder. Apparently, the leap between steps 43 and 44 has the similar nature as that between steps 39 and 40. In this case, the usage of a powerful supercomputer can help inverting 44-step MD4.

\paragraph{Partially constant preimages}
In all found preimages for steps 40 and 43, 9 of 16 message words are equal to $\texttt{0xa57d8668}$. These words were automatically derived because of the known $K$. Recall that $K=\texttt{0xffffffff}$ was used in all cases. However, in some preimages for 41 and 42 steps $M[0]=\texttt{0x257d8668}$, while all remaining 8 message words are equal to $\texttt{0xa57d8668}$. The reason is that in these cases the preimages were found not in the first iteration of Algorithm~\ref{alg:md4-constraints}, so in the 13th step the constant was not $K$, but rather its slightly modified value.

\paragraph{Preimage attacks and second preimage attacks}
In this section and Section~\ref{sec:invert-md4-40}, practical preimage attacks (see Subsection~\ref{subsec:prelim-hash}) on 40-, 41-, 42-, and 43-step MD4 are proposed. Recall that the conquer phase aimed to solve all subproblems (see Section~\ref{sec:alg-cnc}). As a result, 2 preimages were found for the hash \texttt{symm} produced by 40-step MD4, see Table~\ref{tab:runtimes-md4-40}. According to Table~\ref{tab:runtimes-md4-41-42-43}, more that one preimage was found for at least one hash in case of 41-, 42-, and 43-step MD4. Therefore, second preimage attacks are proposed on 40-, 41-, 42-, and 43-step MD4. It is possible that more preimages exist for the considered inverse problems. If an AllSAT solver had been applied to the subproblems instead of a SAT solver, then all preimages would have been found.

%\subsection{Runtime Estimation for Inverting 44-step MD4}\label{subsec:md4-44-estim}

%As it was stated above, adding the 40th step of MD4 gives a leap in hardness of inversion problems, and the next such a leap happens when adding the 44th step. If these leaps have similar nature, than the inversion problems for 44-step MD4 are about 10~000 times harder than that for 40-43 steps, so about 50 years on a single computer are needed to solve one such problem via the same approach. On the other hand, this is reachable on a very powerful supercomputer in just few days. The first natural step is to try more meticulous and time consuming cubing phase. An extremely large value of the parameter $maxc$ of Algorithm~\ref{alg:finding-cutoff}, say 1~000~000~000, can be tried, while $minr$ also should be significantly increased. The estimate should be calculated on the supercomputer. If it is reasonable, then the inversion might be done as well. It also makes sense to try another branching strategies on the cubing phase, see~\cite{Kullmann09}.

%% file: 8.exper-md5.tex
\section{Inverting Unconstrained 28-step MD5}\label{sec:invert-md5}

As mentioned in Subsection~\ref{subsec:prelim-md5}, Dobbertin's constraints are not efficient for MD5. That is why in this study 28-step MD5 is inverted without adding any extra constraints, like it was done in~\cite{LegendreDK12-Ictai}. Recall that in this case for an arbitrary hash there are about $2^{384}$ preimages, but it is not easy to find any of them. Algorithm~\ref{alg:finding-cutoff} in its 
estimating mode is not applicable to MD5 either because the cubing phase gives too hard subproblems for an unconstrained inverse problem, so no runtime estimate can be calculated in reasonable time. On the other hand, since the considered inverse problem has huge number of solutions, the incomplete SAT solving mode of Algorithm~\ref{alg:finding-cutoff} suites well for it.

First a CNF that encodes 28-step MD5 was constructed based on the encoding from Subsection~\ref{subsec:enc-md5}. The CNF has 7~471 variables and 54~672 clauses. The same four hashes were considered for inversion as for MD4: \texttt{0hash}, \texttt{1hash}, \texttt{symmhash}, \texttt{randhash}. Therefore, 4 CNFs were constructed by adding corresponding 128 unit clauses to the original CNF. Then these CNFs were simplified by \textsc{CaDiCaL} such that at most 1 conflict was generated. Characteristics of the simplified CNFs are presented in Table~\ref{tab:cnfs-md5}.

\begin{table}[h]
\caption{Characteristics of the simplified CNFs that encode inverse problems for 28-step MD5.}
\label{tab:cnfs-md5}
\centering
\begin{tabular}{cccc}
\toprule
    Hash & Variables & Clauses & Literals \\
    \midrule
    \texttt{0} & 6~814 & 50~572 & 199~596 \\
    \texttt{1} & 6~844 & 50~749 & 200~153 \\
    \texttt{symm} & 6~842 & 50~737 & 200~114 \\
    \texttt{rand} & 6~842 & 50~741 & 200~110 \\
\bottomrule
\end{tabular}
\end{table}

The SAT solver \textsc{EnCnC} (see the beginning of Subsection~\ref{subsec:exp-setup}) was run on these CNFs in the incomplete SAT solving mode with the following inputs:

\begin{itemize}
\item \textsc{March\_cu}.
\item $\texttt{nstep}=10$.
\item $\texttt{minr} = 0$.
\item $N=1~000$.
\item \textsc{Kissat\_sc2021}.
\item $\texttt{maxst}=5~000$ seconds.
\item $\texttt{cores}=12$.
\item $\texttt{mode}=\texttt{incomplete-solving}$.
\end{itemize}

\begin{figure}[!ht]
   \centering
   \includegraphics[width=.7\textwidth]{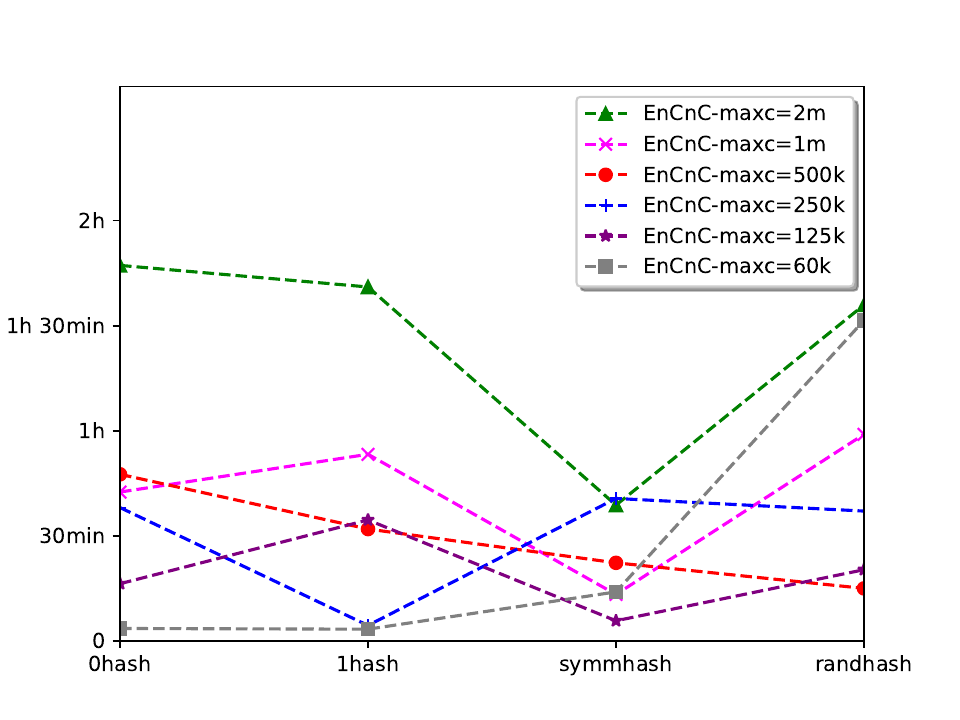}
    \caption{Runtimes of \textsc{EnCnC} in the incomplete SAT solving mode on four MD5-28-related inverse problems.}
    \label{fig:solving-md5-28}
\end{figure}

The key parameter here is $maxc$ (the maximum number of generated cubes), for which the following values were tried: 2~000~000; 1~000~000; 500~000; 250~000; 125~000; 60~000. Note that the default value of $\texttt{maxc}$ in \textsc{EnCnC} is 1~000~000. Recall that in the incomplete SAT solving mode, \textsc{EnCnC} stops if a satisfying assignment is found; if a CDCL solver is interrupted due to a time limit on some subproblem, \textsc{EnCnC} continues working. The corresponding 6 versions of \textsc{EnCnC} with different values of $maxc$ were run on the CNFs with the wall-clock time limit of 1 day. In Table~\ref{tab:solving-md5}, the wall-clock runtimes are presented. Also, the same data is shown in Figure~\ref{fig:solving-md5-28}.

\begin{table}[h]
\caption{Wall clock time for 28-step MD5 on a 12-core CPU. Here ``-'' means that the solver was interrupted due to the time limit of 1 day. The best results are marked with bold.}
\label{tab:solving-md5}
\centering
\begin{tabular}{ccccc}
\toprule
    Solver & \texttt{0hash} & \texttt{1hash} & \texttt{symmhash} & \texttt{randhash}  \\
    \midrule
    \textsc{EnCnC-maxc=2m} & 1 h 47 min & 1 h 41 min & 39 min & 1 h 36 min \\
    \textsc{EnCnC-maxc=1m} & 42 min & 53 min & 13 min & 59 min \\
    \textsc{EnCnC-maxc=500k} & 48 min & 32 min & 22 min & \textbf{15 min} \\
    \textsc{EnCnC-maxc=250k}  & 38 min & 4 min & 41 min & 37 min \\
    \textsc{EnCnC-maxc=125k}  & 16 min & 35 min & \textbf{6 min} & 20 min \\
    \textsc{EnCnC-maxc=60k}  & \textbf{4 min} & \textbf{3 min} & 14 min & 1 h 32 min \\
    % kissat / 12$ & 1145 & 4598 & 601 & 1596 \\
    % kissat on unsimpl CNFs$ & 13734 & 55170 & 7211 & 19151 \\
    % kissat on simpl CNFs & 690 & 31953 & 32097 & 17335 \\
    \midrule
    \textsc{P-MCOMSPS} & - & - & - & - \\
    \midrule
    \textsc{Treengeling} & - & - & - & - \\
\bottomrule
\end{tabular}
\end{table}

In addition, two complete parallel CDCL SAT solvers were tried. The first one, \textsc{P-MCOMSPS}, is the winner of the Parallel track in SAT Competition 2021~\cite{ValladeFOBSKNG21-Pmcomsps}. It is a portfolio solver built upon the widely-used \textsc{Painless} framework~\cite{FriouxBSK17-Painless}. The second one, \textsc{treengeling}~\cite{Biere16-Treeng}, is a Cube-and-Conquer solver. It was chosen to compare \textsc{EnCnC} with a competitor built upon a similar strategy. Besides this, \textsc{treengeling} won several prizes in SAT Competitions and SAT Races.

Let us discuss the results. Based on average runtime, the best version of \textsc{EnCnC} is \textsc{EnCnC-maxc=125k}, while the worst is \textsc{EnCnC-maxc=2m}. Nevertheless, all versions managed to find satisfying assignments for all 4 CNFs within the time limit. At 23 runs out of 24, versions of \textsc{EnCnC} did it during solving the first 12 subproblems from the first random sample (for the lowest values of the cutoff threshold). It means that \textsc{Kissat} did not reach the time limit of 5~000 seconds in these cases. The only exception is \textsc{EnCnC-maxc=60k} on \texttt{randhash}, where on all 12 first subproblems \textsc{Kissat} was interrupted due to the time limit, and then a satisfying assignment was found in one of the next 12 subproblems from the same sample. As for the competitors, they could not solve anything within the time limit. In Table~\ref{tab:preimages-md5-28}, the preimages found by \textsc{EnCnC-maxc=2m} are presented. It should be noted that preimages for \texttt{0hash}, \texttt{1hash}, and \texttt{randhash} have not been published so far.

\begin{table}
{\caption{Preimages found by \textsc{EnCnC-maxc=2m} for 28-step MD5.}
\label{tab:preimages-md5-28}}
\centering
\begin{tabular}{cl}
\toprule
    Hash & Preimages \\
    \midrule
    \multirow{2}{*}{\texttt{0}}
    & \scriptsize{0xd825e4fb 0xa73fcaa9 0x660cd53d 0xb9308515 0x4677d4e0 0xcadcee62 0x40722cb3 0xf41a4b12} \\ 
	& \scriptsize{0xac2fdec3 0x9cbcb4a3 0xffcca30f 0x9a0e2026 0x475763e5 0x30ce233b 0xbef0cd57 0x1a6b39d} \\ 
    \midrule
    \multirow{2}{*}{\texttt{1}} 
	  & \scriptsize{0xdfe6feeb 0xc4437a85 0x11af5182 0xe3b13f03 0x5103e1fc 0xea231da2 0xc3b513d1 0xb95fa9d7} \\
	& \scriptsize{0x7a2a331c 0x2ddf2607 0x699a2dae 0xc1827561 0xfe80aeed 0xcf45b09a 0x5b596c8f 0xd0265347} \\
    \midrule
    \multirow{2}{*}{\texttt{symm}}
    & \scriptsize{0x54032182 0x2a1693f1 0x1053aef3 0x9f4d7c87 0x9f0d5ba1 0xb43a63f8 0x4310aa89 0x9df4e0d8} \\
    & \scriptsize{0xada73cbf 0x63fd55c2 0x49f1f4a0 0x5e05beff 0x6c149122 0x54a25f8e 0x12ef4bb0 0x78482fb4} \\
    \midrule
    \multirow{2}{*}{\texttt{rand}}
    & \scriptsize{0x120686db 0xad5834c6 0x7d660963 0x71c408fe 0x17cf4511 0x75df78de 0x544ae232 0x13745ecc} \\
    & \scriptsize{0x9190f8a2 0x4878ab8d 0x43229cc7 0x5013f2de 0xd49b395a 0xa151b704 0x5f1dd4ec 0xc860dfb5} \\
\bottomrule
\end{tabular}
\end{table}

The found preimages were verified by the reference implementation from~\cite{Rivest92-MD5}. It can be easily reproduced in the same way that was discussed in Subsection~\ref{subsec:discussion}.

%% file: 9.related.tex
\section{Related Work}\label{sec:related}

% SAT-based cryptanalysis
Apparently, SAT-based cryptanalysis was first proposed in 1996~\cite{CookM96-SATcrypt}, but for the first time it was applied to solve a real cryptanalysis problem in 2000~\cite{MassacciM00-LogCrypt}. In particular, a reduced version of the block cipher DES was analyzed there via a SAT solver. Since that publication, SAT-based cryptanalysis has been successfully applied to analyze various block ciphers, stream ciphers, and cryptographic hash functions~\cite{Bard09-AlgCrypt}.

% SAT-based cryptanalysis of cyrptographic hash functions
SAT-based cryptanalysis via CDCL solvers has been applied earlier to cryptographic hash functions of the MD family as follows. For the first time it was done in~\cite{JovanovicJ05-hash} to construct benchmarks with adjustable hardness. In~\cite{MironovZ06-hash}, a practical collision attack on MD4 was performed. 39-step MD4 was inverted in~\cite{DeKV07-MD4,LegendreDK12-Ictai,LafitteNH14-MD4,GribanovaZKOS17-MitMD,GribanovaS18-MD4}. In~\cite{GladushGKPS22-Pavt}, the hardness of practical preimage attacks on 43-, 45-, and 47-step MD4 was estimated. In \cite{GribanovaS20-WeakMD4}, an MD4-based function was constructed and the full (48-step) version of this function was inverted. As for MD5, in~\cite{MironovZ06-hash} and later in~\cite{GribanovaZKOS17-MitMD}, practical collision attacks on MD5 were performed. In~\cite{DeKV07-MD4}, 26-step MD5 was inverted, while in \cite{LegendreDK12-Ictai} it was done for 27- and 28-step MD5. 

% SHA-1
Also, CDCL solvers were applied to analyze cryptographic hash functions from the SHA family. Note that SHA-1 is an improved version of MD4. For the first time a collision for SHA-1 was found in~\cite{StevensBKAM17-SHA1collision} and it was done partially by a CDCL solver. Step-reduced versions of SHA-0, SHA-1, SHA-256, and SHA-3 were inverted in~\cite{Nossum2012,LegendreDK12-Ictai,HomsirikamolMRS12-SHA3,NejatiLGCG17-SHA1}. An algebraic fault attack on SHA-1 and SHA-2 was performed in~\cite{NejatiHGG18-SHA}, while that on SHA-256 was done in~\cite{NakamuraHH21-AFAsha256}.

The first theoretical preimage attack on MD4 with the complexity of $2^{102}$ was proposed in~\cite{Leurent08-MD4}. Later the complexity was reduced to $2^{99.7}$~\cite{GuoLRW10-TheorAttackMD4}. As for MD5, the best theoretical attack has the complexity of $2^{123.4}$~\cite{SasakiA09-md5}.

% Non-cryptographic problems solved by CnC:
The following hard mathematical problems have been solved via Cube-and-Conquer: the Erd{\H{o}}s discrepancy problem~\cite{KonevL15-Erdos}; the Boolean Pythagorean Triples problem~\cite{HeuleKM16-BPT}; Schur Number Five~\cite{Heule18-Schur5}; Lam's problem~\cite{BrightCSKG21-Lam}; Keller’s Conjecture~\cite{BrakensiekHMN22-KellerConj}. In~\cite{LiBG2024-KochenSpecker}, the lower bound for the Minimum Kochen–Specker Problem was improved. In~\cite{WeaverH20-PerfHashFunc}, new minimal perfect hash functions were found. Note that these hash functions are not cryptographic ones and find their application in lookup tables. In the present paper, for the first time significant cryptanalysis problems were solved via Cube-and-Conquer.

% Estimating the hardness of SAT instances:
The present paper presents a general Cube-and-Conquer-based algorithm for estimating hardness of SAT instances. Usually this is done by other approaches: the tree-like space complexity~\cite{AnsoteguiBLM08-measure}; supervised machine learning~\cite{HutterXHL14-AIJ}; the popularity–similarity model~\cite{Almagro-BlancoG22-SatTemper}; backdoors~\cite{WilliamsGS03-Backdoor}.

% Connection with backdoors:
Backdoors are closely connected with Cube-and-Conquer. Informally, backdoor is a subset of variables of a given formula, such that by varying all possible values of the backdoor's variables simpler subproblems are obtained which can be solved independently~\cite{WilliamsGS03-Backdoor,KilbySTW05-Backdoors,DilkinaGS07-Backdoors,SamerS08-BackdoorTrees}. In fact, a set of backdoor's values can be considered a cube, while choosing a proper backdoor and varying all corresponding values is a special way to generate cubes in the cubing phase of Cube-and-Conquer. For a given SAT instance and a backdoor, hardness of the instance can be estimated by processing a (relatively small) sample of subproblems~\cite{SemenovZBP11-PactA5-1}. 

The search for a backdoor with the minimum hardness was reduced to minimization of pseudo-Boolean objective functions in application to SAT-based cryptanalysis in~\cite{SemenovZOKI18-IBS,KochemazovZ18-Alias,SemenovCPOUI21-BackdoorsCP}. In~\cite{ZaikinK21-OMS,SemenovZK2021-handbook} it was shown that these functions are costly, stochastic, and black-box. In the present paper, a pseudo-Boolean objective function with the same properties is minimized to find a cutoff threshold of the cubing phase of Cube-and-Conquer rather than a backdoor.

%% file: 10.conc.tex
\section{Conclusions and Future Work}\label{sec:concl}

This paper proposed two algorithms. Given a hash, the first algorithm gradually modifies one of twelve Dobbertin's constraints for MD4 until a preimage for a given hash is found. Any complete algorithm can be used to solve the corresponding intermediate inverse problems. The second proposed algorithm can operate with a given CNF in two modes. In the estimating mode, cutoff thresholds of the cubing phase of Cube-and-Conquer are varied, and the CNF's hardness for each threshold is estimated via sampling. The threshold with the best estimate can be naturally used to choose a proper computational platform and solve the SAT instance if the estimate is reasonable. This mode is general, so it can be applied to estimate the hardness and solve hard SAT instances from various classes. In the incomplete SAT solving mode, the second algorithm is a SAT solver oriented on satisfiable CNFs with many satisfying assignments. 

The cryptographic hash function MD4 was analyzed by a combination of the first algorithm and the estimating mode of the second algorithm, which were implemented as a multithreaded program. As a result, the first practical preimage attacks and second preimage attacks on 40-, 41-, 42-, and 43-step MD4 were performed on a computer. In contrast to MD4, MD5 served as an example of a cryptographic hash function for which no problem-specific constraints were added. By applying the incomplete SAT solving mode of the second algorithm, preimages for two most regular hashes (128 1s and 128 0s) produced by 28-step MD5 were found for the first time.

In the future we plan to apply the proposed algorithms to analyze other cryptographic hash functions: SHA-1, SHA-2, RIPEMD. Also we are going to investigate two MD4-related phenomena which were figured out during the experiments. The first one is the non-effectiveness (in most cases) of an advanced simplification in application to the constructed CNFs. The second one is an evident division of subproblems in the conquer phase to extremely simple ones and hard ones. Finally, we plan to compare the estimating mode of the second proposed algorithm with other approaches, which are usually used to estimate the hardness of a given SAT instance.

%% file: appendix.tex
\appendix

\section{Estimates for Step-reduced MD4}\label{appendix:figures}

The following figures depict how the objective function was minimized on 40- and 43-step MD4.

\begin{figure}[ht]
\centering
\subfloat[\emph{MD4-40, \emph{1hash}}]{\label{fig:est-md4-40-1hash}\includegraphics[scale=.48]{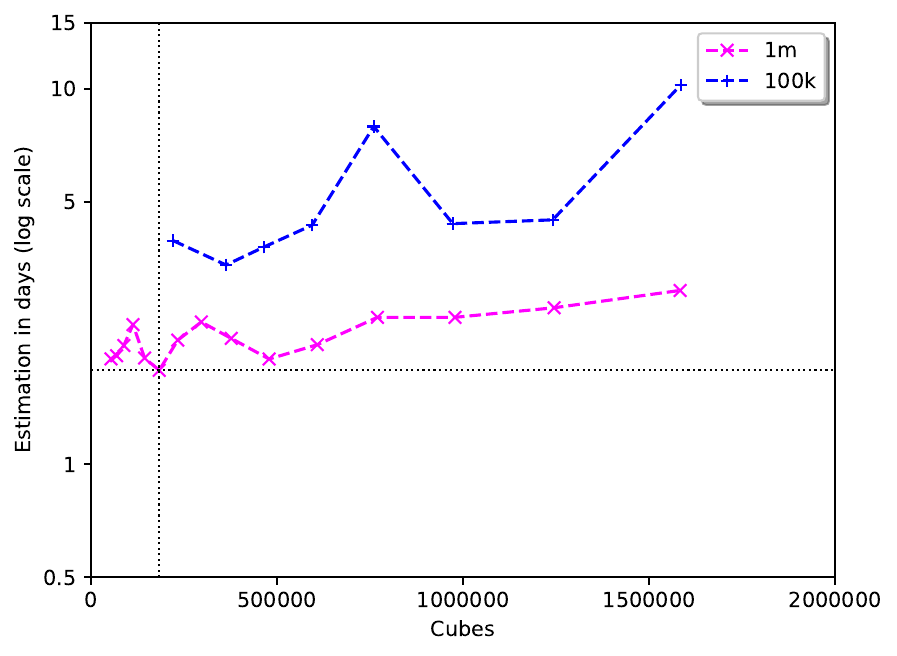}}\quad
\subfloat[\emph{MD4-43, \emph{0hash}}]
{\label{fig:est-md4-43-0hash}\includegraphics[scale=.48]{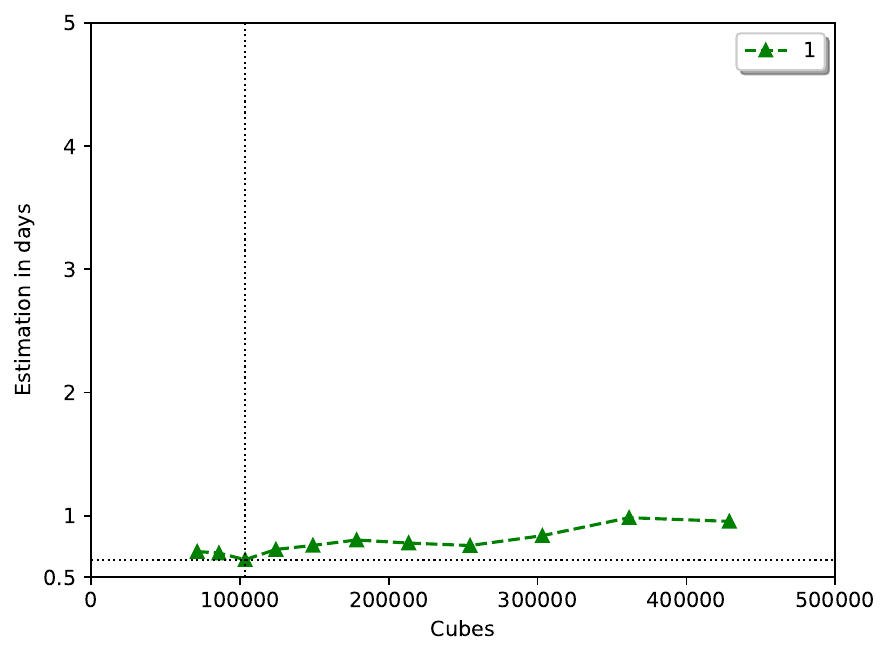}}\quad
\subfloat[\emph{MD4-40, \emph{symmhash}}]
{\label{fig:est-md4-40-symmhash}\includegraphics[scale=.48]{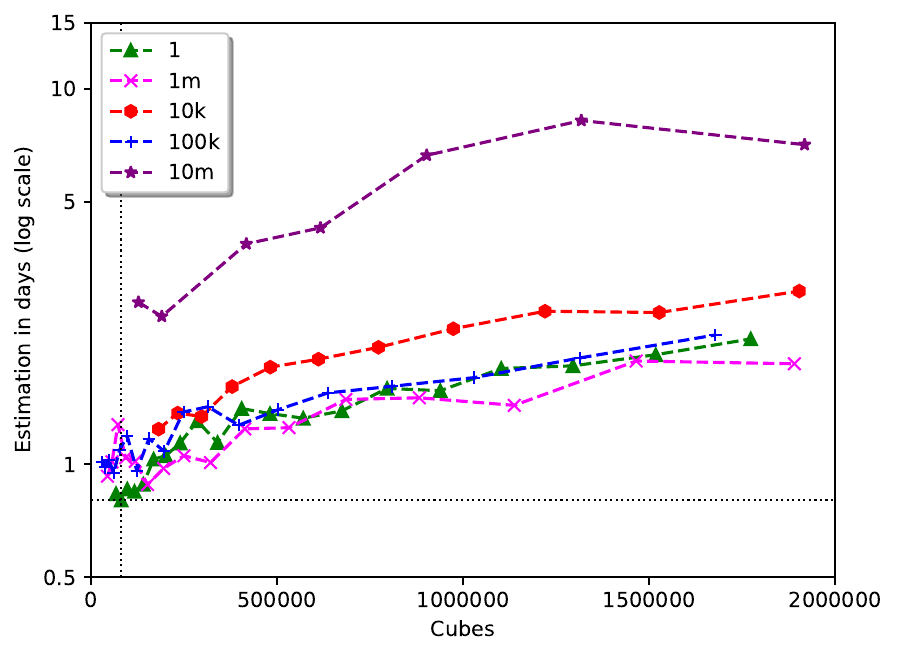}}
\subfloat[\emph{MD4-43, \emph{symmhash}}]
{\label{fig:est-md4-43-symmhash}\includegraphics[scale=.48]{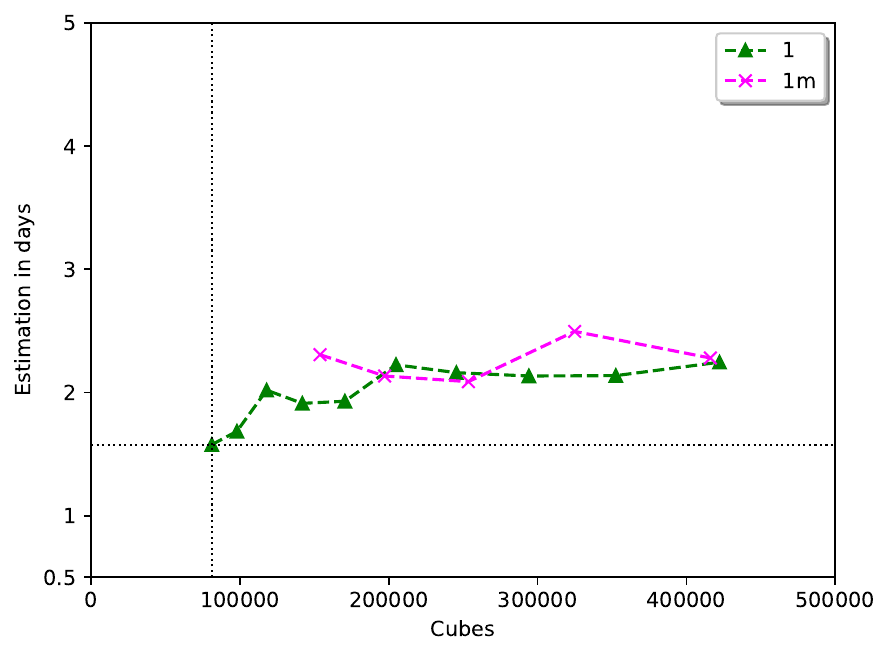}}\quad
\subfloat[\emph{MD4-40, \emph{randhash}}]
{\label{fig:est-md4-40-randhash}\includegraphics[scale=.48]{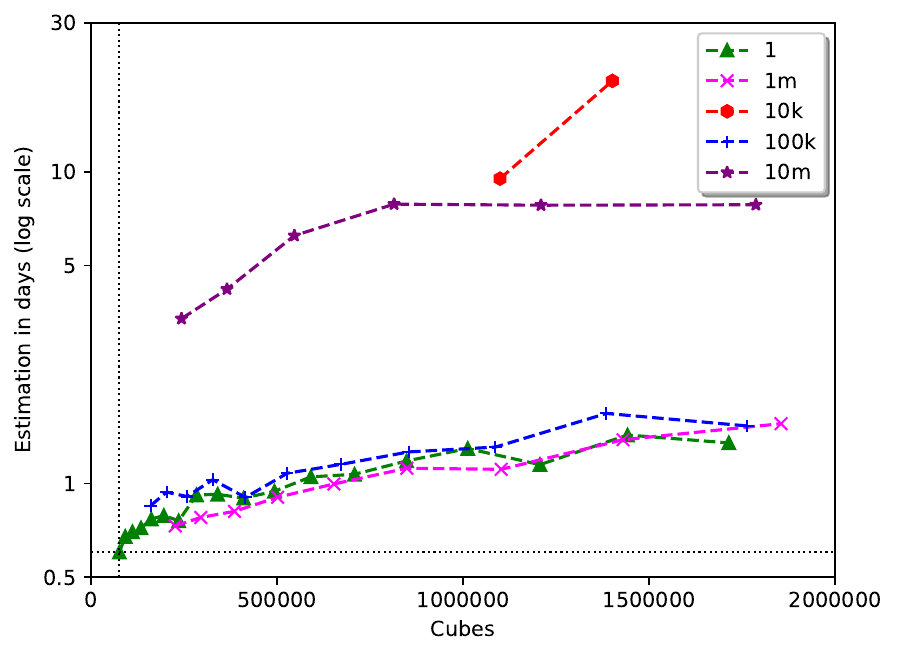}}
\subfloat[MD4-43, \emph{randhash}]
{\label{fig:est-md4-43-randhash}\includegraphics[scale=.48]{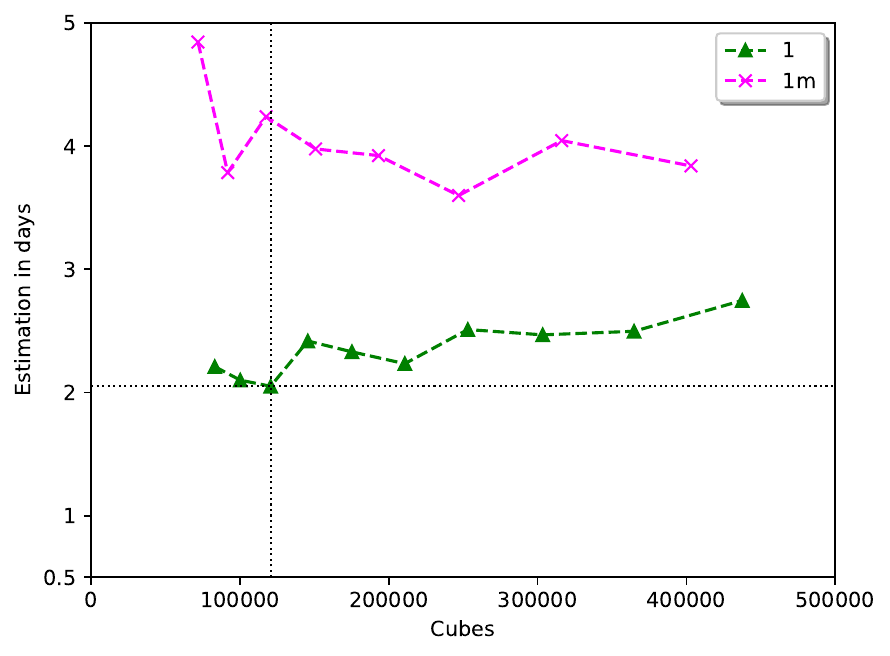}}
\caption{Minimization of the objective function on 40- and 43-step MD4.}
\label{fig:est-md4-40-43}
\end{figure}

\section{Found Preimages for Step-reduced MD4}\label{appendix:preimages}

\begin{table}[!ht]
{\caption{Found preimages for 41-step MD4.}
\label{tab:preimages-md4-41}}
\centering
\begin{tabular}{cl}
\toprule
    %Hash & Preimages \\
    %\hline
    \multirow{2}{*}{\emph{0}}
    & \scriptsize{0x257d8668 0xa57d8668 0xa57d8668 0xdafb914d 0xa57d8668 0xa57d8668 0xa57d8668 0x1edf9f78} \\
    & \scriptsize{0xa57d8668 0xa57d8668 0xa57d8668 0x12984195 0x97f0b6c 0xd9e5df17 0xabe482c7 0x23d98522} \\
    \midrule
    \multirow{2}{*}{\emph{1}} 
	  & \scriptsize{0xa57d8668 0xa57d8668 0xa57d8668 0x5c31dc3 0xa57d8668 0xa57d8668 0xa57d8668 0x52f59fb2} \\
    & \scriptsize{0xa57d8668 0xa57d8668 0xa57d8668 0x1e8a7cbb 0x3982e99f 0x812d980d 0x27b8d0b5 0xb81a00d1} \\
    \midrule
    \multirow{8}{*}{\emph{symm}}
    & \scriptsize{0x257d8668 0xa57d8668 0xa57d8668 0xeaaf86e 0xa57d8668 0xa57d8668 0xa57d8668 0xc3b97274} \\
    & \scriptsize{0xa57d8668 0xa57d8668 0xa57d8668 0x21b8d189 0x15fc5540 0xd283c2c4 0x7d27396b 0x7bb74632} \\
    \cline{2-2}
    & \scriptsize{0x257d8668 0xa57d8668 0xa57d8668 0x5e8d818a 0xa57d8668 0xa57d8668 0xa57d8668 0x8fc29cce} \\
    & \scriptsize{0xa57d8668 0xa57d8668 0xa57d8668 0x8c6b49cc 0xe31a2c8d 0x9a5e1c5d 0x2dd896f5 0x1ed72fab} \\
    \cline{2-2}
    & \scriptsize{0x257d8668 0xa57d8668 0xa57d8668 0x9278c8f 0xa57d8668 0xa57d8668 0xa57d8668 0x4e3194eb} \\
    & \scriptsize{0xa57d8668 0xa57d8668 0xa57d8668 0x22efb603 0xe2b4a054 0xd74ec43 0xf09b0821 0xe4ca9fca} \\
    \cline{2-2}
    & \scriptsize{0x257d8668 0xa57d8668 0xa57d8668 0x6172bd01 0xa57d8668 0xa57d8668 0xa57d8668 0x8e35540f} \\
    & \scriptsize{0xa57d8668 0xa57d8668 0xa57d8668 0x4b8210a9 0xd5c0fedb 0x45c28d93 0x1b542bb8 0x74c28676} \\
    \midrule
    \multirow{6}{*}{\emph{rand}} 
    & \scriptsize{0xa57d8668 0xa57d8668 0xa57d8668 0x4b11d0ca 0xa57d8668 0xa57d8668 0xa57d8668 0x4c195670} \\
    & \scriptsize{0xa57d8668 0xa57d8668 0xa57d8668 0x76529071 0x68d3862d 0xdd3779df 0x768ce847 0x77e1b04e} \\
    \cline{2-2}
    & \scriptsize{0xa57d8668 0xa57d8668 0xa57d8668 0xcfbf3444 0xa57d8668 0xa57d8668 0xa57d8668 0xaac69f2f} \\
    & \scriptsize{0xa57d8668 0xa57d8668 0xa57d8668 0xbdaf1de9 0xfb9496dc 0x537e7a8c 0xd083975f 0xf3a5fc76} \\
    \cline{2-2}
    & \scriptsize{0xa57d8668 0xa57d8668 0xa57d8668 0xbfbf37eb 0xa57d8668 0xa57d8668 0xa57d8668 0xf3252a5c} \\
    & \scriptsize{0xa57d8668 0xa57d8668 0xa57d8668 0x3f829fe3 0x28c0fe6 0x27eadfa1 0xc87af86e 0x48fcd23d} \\
\bottomrule
\end{tabular}
\end{table}

\begin{table}[!ht]
{\caption{Found preimages for 42-step MD4.}
\label{tab:preimages-md4-42}}
\centering
\begin{tabular}{cl}
\toprule
    %Hash & Preimages \\
    %\hline
    \multirow{6}{*}{\emph{0}}
    & \scriptsize{0xa57d8668 0xa57d8668 0xa57d8668 0xecdab667 0xa57d8668 0xa57d8668 0xa57d8668 0xe3844a01} \\
    & \scriptsize{0xa57d8668 0xa57d8668 0xa57d8668 0xa3205929 0xfad1ea59 0xd2cae4d2 0x52149d55 0xc82cffbf} \\
    \cline{2-2}
    & \scriptsize{0xa57d8668 0xa57d8668 0xa57d8668 0xae60af85 0xa57d8668 0xa57d8668 0xa57d8668 0x8bcd69e3} \\
    & \scriptsize{0xa57d8668 0xa57d8668 0xa57d8668 0x59b8bf6 0x7755a76 0xfbe0b515 0xf9a31765 0x14d516a6} \\
    \cline{2-2}
    & \scriptsize{0xa57d8668 0xa57d8668 0xa57d8668 0xa9210d09 0xa57d8668 0xa57d8668 0xa57d8668 0xba9694ea} \\
    & \scriptsize{0xa57d8668 0xa57d8668 0xa57d8668 0x6a8157fe 0xd6566aae 0xbacb3d6c 0x1ec4854d 0x22357d65} \\
    \midrule
    \multirow{2}{*}{\emph{1}} 
	  & \scriptsize{0x257d8668 0xa57d8668 0xa57d8668 0xd8f77148 0xa57d8668 0xa57d8668 0xa57d8668 0x88275d15} \\
    & \scriptsize{0xa57d8668 0xa57d8668 0xa57d8668 0xcf6b92d0 0x4a8e498d 0x3beb0878 0xb55e027 0x87b4d62c} \\
    \midrule
    \multirow{2}{*}{\emph{symm}}
    & \scriptsize{0xa57d8668 0xa57d8668 0xa57d8668 0xd1dce7ea 0xa57d8668 0xa57d8668 0xa57d8668 0xcbc2a90} \\
    & \scriptsize{0xa57d8668 0xa57d8668 0xa57d8668 0xd9834f6d 0x5267d5d6 0x41a9cf18 0x71469663 0xbd507731} \\
    \midrule
    \multirow{6}{*}{\emph{rand}} 
    & \scriptsize{0x257d8668 0xa57d8668 0xa57d8668 0xbd7389e6 0xa57d8668 0xa57d8668 0xa57d8668 0x3eb8ae3a} \\
    & \scriptsize{0xa57d8668 0xa57d8668 0xa57d8668 0x162c323e 0xa4056a04 0x9da74aac 0xfee2c77 0x8b25de8e} \\
    \cline{2-2}
    & \scriptsize{0x257d8668 0xa57d8668 0xa57d8668 0xc1748842 0xa57d8668 0xa57d8668 0xa57d8668 0xd7e32a57} \\
    & \scriptsize{0xa57d8668 0xa57d8668 0xa57d8668 0x21c5baab 0x552a7372 0xa21b2963 0x2fe88ffb 0xadfddb3} \\
    \cline{2-2}
    & \scriptsize{0x257d8668 0xa57d8668 0xa57d8668 0xc455558f 0xa57d8668 0xa57d8668 0xa57d8668 0xff87976a} \\
    & \scriptsize{0xa57d8668 0xa57d8668 0xa57d8668 0x3e82e858 0x46ad9cde 0x76f3b1d0 0x31aadb79 0x45cc1c91} \\
\bottomrule
\end{tabular}
\end{table}